\documentclass[twocolumn,aps,showpacs,amsmath,amssymb]{revtex4}

\usepackage{graphicx}
\usepackage{dcolumn}
\usepackage{bm}

\begin{document}

To be published in Chinese Physics C.

\title{
Production of super-heavy nuclei in cold fusion reactions
}%

\author{
V. Yu. Denisov$^{1,2;}$\footnote{denisov@kinr.kiev.ua}%
\quad I. Yu. Sedykh$^{3}$,
}%

\affiliation{%
$^{1}$ Institute for Nuclear Research, Prospect Nauki 47,
03028 Kiev, Ukraine \\
$^{2}$ Faculty of Physics, Taras Shevchenko National University of Kiev, Prospect Glushkova 2, 03022 Kiev, Ukraine \\
$^{3}$ Financial University under the Government of the Russian Federation, Leningradsky Prospekt 49, 125993 Moscow, Russian Federation
}%

\begin{abstract}
The model for the cold-fusion reactions related to the synthesis of super-heavy nuclei in collisions of heavy projectile-nuclei with $^{208}$Pb target nucleus is discussed. In the framework of this model the production of the compound nucleus by two paths through, the di-nuclear system and the fusion way, are taken into account simultaneously. The formation of the compound nucleus in the framework of the di-nuclear system is related to the transfer of nucleons from the light nucleus to the heavy one. The fusion way is linked to the sequential evolution of the nuclear shape from the system of contacting nuclei to the compound nucleus. It is shown that the compound nucleus is mainly formed by the fusion way in the cold-fusion reactions. The landscape of the potential energy related to the fusion path is discussed in detail. This landscape for very heavy nucleus-nucleus systems has the intermediate state, which is linked to the formation of both the compound nucleus and the quasi-fission fragments. The decay of the intermediate state is taken into account in the calculation of the compound nucleus production cross sections and the quasi-fission cross sections. The values of the cold-fusion cross sections obtained in the model are well agreed with the experimental data.
\end{abstract}

\maketitle

\section{introduction}

Synthesis of Super-Heavy nuclei (SHN) is very interesting, exciting and puzzling physical task as for experimentalists as for theoreticians. The elements beyond Md with the number of protons $Z= 102-118$ were synthesized by the fusion of heavy nuclei \cite{h,h-b,h-c,102jinr,102jinr1,104gsi,104gsi-a,104lbnl,106gsi,106lbnl,108gsi,108riken,110riken,110lbnl,112gsi,112riken,112riken1,folden,m,m1,m3,og,og1,118,118a,sh,dp}. The element Og with $Z=118$ is the heaviest element, which has been synthesized to now by Oganessian et al. \cite{og,118}. Recently, the experiments aimed at the synthesis of isotopes of element Z = 119 and 120 or study the properties of related reactions have been performed \cite{119,120,120a,120b,120c,120d,120e}, but no decay chains consistent with the fusion-evaporation reaction products were observed.

The cold-fusion reactions for SHN synthesis are the reactions between heavy-ion projectiles with the mass/charge $A \geq 48 / Z \geq 20$ and lead or bismuth targets. This type of reaction has been proposed by Oganessian et al. \cite{cfr}. Using these reactions the SHN with charges Z=102$\div$113 have been successfully synthesized in experiments \cite{h,h-b,h-c,102jinr,102jinr1,104gsi,104gsi-a,104lbnl,106gsi,106lbnl,108gsi,108riken,110riken,110lbnl,112gsi,112riken,112riken1,folden,m,m1,m3}.

Many different models have been proposed for a description of the cross section of the synthesis of the SHN in heavy-ion collisions, see, for example, Refs. \cite{dp,sw,sw1,sw2,sw3,sww2,sww3,sww4,dh,d-sym,zg,jap1,jap4,shen1,kos2,kos3,boilley,abe3,liang1,umar,umar1,umar2,torres2,torres,torres1,v,dns1,dns1c,dns1aa,dns1aa1,dns1aa2,dns1aa4,dns2a,adamian,dns3,dns31,dns32,dns33,dns3a,dns4b,dns4c,dns4d,dns5,dns444,dns4444,param,param1c,param1d,param2c,param3,param4a,param4b,param5,param6,param7} and papers cited therein. The common feature of the models is that the cross section of the SHN production is described as the product of the capture cross section, the probability of compound nucleus formation, and the survival probability of the compound nucleus. The capture process is related to the formation of the system of touching nuclei. The probability of compound nucleus formation is connected to the process of the evolution from the system of contacting nuclei to the spherical or near spherical compound nucleus. The survival probability of compound nucleus is linked to the competition between the fission and neutron emission processes. When the excitation energy of a compound nucleus drops down the residue nuclei emits alpha-particles and/or divided on two fission fragments, which are detected in experiments. The chain of alpha-particles and/or the fission fragments observed in the experiment are the experimental signals of the successful synthesis of the SHN \cite{h,h-b,h-c,102jinr,102jinr1,104gsi,104gsi-a,104lbnl,106gsi,106lbnl,108gsi,108riken,110riken,110lbnl,112gsi,112riken,112riken1,folden,m,m1,m3,og,og1,118,118a,sh,dp}.

The differences between various models \cite{sw,sw1,sw2,sw3,sww2,sww3,sww4,dh,d-sym,zg,jap1,jap4,shen1,kos2,kos3,boilley,abe3,liang1,umar,umar1,umar2,torres1,torres2,torres,v,dns1,dns1c,dns1aa,dns1aa1,dns1aa2,dns1aa4,dns2a,adamian,dns3,dns31,dns32,dns33,dns3a,dns4b,dns4c,dns4d,dns5,dns444,dns4444,param,param1c,param1d,param2c,param3,param4a,param4b,param5,param6,param7} at the description of the capture process are related to using of the different nucleus-nucleus interaction potentials, consideration of different shapes and mutual orientations of interacting nuclei, and applications of various approximations at the evaluation of the capture cross sections. The differences of various models at the description of the survival probability are connected to applying of different statistical nuclear decay models, various expressions for the energy level densities, different values of the fission barrier of the SHN obtained in the frameworks of various nuclear structure models, various dependencies of the fission barrier on the excitation energy of the compound nucleus, different values of the neutron binding energy taken from various nuclear mass models. Consequently, the corresponding differences between the results obtained in the frameworks of various approaches to the capture cross-section and the survival probability of the compound nucleus are natural, reasonable and understandable. The physical mechanisms of the capture cross-section and the survival probability of the compound nucleus are the same in various models.

The process of the compound nucleus formation from the two touching nuclei during the synthesis of SHN is most undefined up to now, because there are the two alternative mechanisms of this process, the fusing \cite{sw,sw1,sw2,sw3,sww2,sww3,sww4,dh,d-sym,zg,jap1,jap4,shen1,kos2,kos3,boilley,abe3,liang1,torres2,torres,umar,umar1,umar2} and Di-Nuclear System (DNS) \cite{v,dns1,dns1c,dns1aa,dns1aa1,dns1aa2,dns1aa4,dns2a,dns3,dns31,dns32,dns33,dns3a,dns4b,dns4c,dns4d,adamian,dns5,dns444,dns444,}, which are actively discussing.

The fusing approach to the compound nucleus formation is related to the smooth subsequent shape evolution from the system of two touching nuclei to the compound nucleus, see Refs. \cite{sw,sw1,sw2,sw3,sww2,sww3,sww4,dh,d-sym,zg,jap1,jap4,shen1,kos2,kos3,boilley,abe3,liang1,umar,umar1,umar2,torres,torres2} and papers cited therein. The models considered in Refs. \cite{sw,sw1,sw2,sw3,sww2,sww3,sww4,dh,d-sym,zg,jap1,jap4,shen1,kos2,kos3,boilley,abe3,liang1,umar,umar1,umar2,torres,torres2} are based on different approaches to the shape evolution and/or the shape parametrization of the nuclear system during the compound nucleus formation. The compound nucleus formation is formed in a competition to the quasi-fission and deep-inelastic processes in the framework of the fusing approach.

The alternative mechanism of the compound nucleus formation from the two colliding nuclei is proposed by Volkov in Ref. \cite{v}. This mechanism is related to the evolution of DNS formed by the two contacting nuclei after penetration through the fusion barrier of the incident nuclei. In the framework of the DNS model the compound nucleus is formed by the multi-nucleon transfer from the light nucleus to the heavy one. The transfer of nucleons in the opposite direction (from heavy nucleus to the light one) as well as the decay of the DNS through the fusion barrier are the processes competing to the compound nucleus formation. The both nuclei touch each other during these multi-nucleon transfers. This DNS mechanism of the compound nucleus formation is applied to the SHN production in Refs. \cite{v,dns1,dns1c,dns1aa,dns1aa1,dns1aa2,dns1aa4,dns2a,dns3,dns31,dns32,dns33,dns3a,dns4b,dns4c,dns4d,adamian,dns5,dns444,dns4444}, see also papers cited therein.

Note, the same experimental of values of the production cross section of SHN are described by applying various approaches for the compound nucleus formation. However, the fusion and DNS mechanisms of compound nucleus formation are very different. It should be also noted that the probability of the compound nucleus formation is also evaluated by various phenomenological or semi-phenomenological expressions \cite{sw,sw1,sw2,sww2,sww3,param,param1c,param1d,param2c,param3,param4a,param4b,param5,param6,param7}.

For the first time the DNS approach is successfully applied to a description of the deep inelastic heavy-ion collisions \cite{v2}. In this case, two nuclei collide at high energies and form a fast-rotating DNS. The fast rotation of DNS prevents the fusion of two nuclei and stabilize the rotating DNS, because the potential of rotating DNS shows the repulsion at small distances between nuclei due to the contribution of the centrifugal force. Therefore, the multi-nucleon exchange between the fast-rotating nuclei forming the DNS is naturally taken place during the deep inelastic heavy-ion collisions. After rotation on some angle the DNS decays on two excited nuclei, which have other nucleon compositions and smaller values of the relative kinetic energies than the incident nuclei \cite{v2}.

The cold fusion reactions are taking place at the collision energies near the barrier of the nucleus-nucleus potential \cite{dn,dnest}. Due to this the possible frequency of rotation of the DNS formed in the SHN formation reaction is much smaller than the one in the case of the deep inelastic heavy-ion collisions. Therefore, the stabilization of DNS due to rotation is impossible for heavy-ion systems leading to the SHN. Another possibility of the of the formation of the barrier at small distances between ions is a diabatic behaviour of nuclear levels during fast collisions \cite{blnr}. The diabatic shift of heavy-ion potential energy occurs when the relative velocity of heavy ions is very high, therefore nucleons occupy diabatic levels and cannot quickly relax in time to the adiabatic ones. However, the relative velocity of ions is disappeared during the penetration of the fusion barrier and the evolution of dinuclear system is related to small relative velocities. The potential energy surfaces obtained in the framework of various microscopic or semimicroscopic calculations have not show the barrier for reactions related to the SHN synthesis \cite{zg,jap1,jap4,shen1,kos2,kos3,boilley,abe3,umar,umar1,umar2}. Nevertheless, the DNS potential calculated in Refs. \cite{dns1,dns1c,dns1aa,dns1aa1,dns1aa2,dns1aa4,dns2a,adamian,dns3,dns31,dns32,dns33,dns3a,dns4b,dns4c,dns4d,dns5,dns444,dns4444} shows the strong repulsion at small distances between nuclei, which stabilizes the DNS system in this model. Such behaviour of the DNS potential in Refs. \cite{dns1,dns1c,dns1aa,dns1aa1,dns1aa2,dns1aa4,dns2a,adamian,dns3,dns31,dns32,dns33,dns3a,dns4b,dns4c,dns4d,dns5,dns444,dns4444} relates to the calculation of the nucleus-nucleus potential in the frozen-density approach. The frozen-density nucleus-nucleus potential is strongly repulsive at small distances between nuclei, because the frozen nucleon densities of the colliding nuclei overlap well at such distances and form a high-density region \cite{dn,dnest}. Due to the high value of the incompressibility of nuclear matter the potential energy of nucleus-nucleus system drastically rises, when the nucleon density became higher than the equilibrium density of nuclear matter \cite{dn,dnest}.

The density distribution of nucleons in colliding nuclei can relax during collisions at small relative velocities of nuclei, which are taken place at the collision energies close the barrier. Therefore, the high nucleon density region with the density noticeably higher than the equilibrium density of nuclear matter is not formed during heavy-ion collision leading to the SHN. As a result, the realistic nucleus-nucleus potential does not show strong repulsion at small distances between two nuclei. The behaviour of the potential at such distances is related to the sequential evolution of the shape of nuclear system. Therefore, the nuclei can fuse and the both mechanisms of compound nucleus formation, the fusion and the DNS, should be simultaneously taken into account in the calculation of the SHN production in heavy-ion collisions. The both mechanisms of the compound nucleus formation are shortly considered in Ref. \cite{torres1}. Below we discuss new model for SHN formation, which includes the both mechanisms of the compound nucleus formation and also the decay of primary-formed excited compound nucleus related to the neutron evaporation in competition with fission. We propose new shape parametrization for a description the fusion path, obtain the cross section of SHN in cold fusion reactions, and compare the calculated cross section values with the experimental data.

Below we present the new model for the description of the cross section of the SHN synthesis in heavy-ion collisions, which takes into account simultaneously both the fusion and DNS mechanisms of the compound nucleus formation. The mechanisms of the nucleus-nucleus capture and the survival of the compound nucleus applied in our model are close to tradition the ones, but we introduce some new features at the consideration of the capture and survival stages of SHN synthesis. The detail description of our model is presented in Sec. II. The discussion of results obtained in our model for the cold fusion reactions is given in Sec. III. Conclusions are drawn in Sec. IV.

\section{The model}

The cold fusion reactions of the SHN synthesis are related to the collisions of the projectiles nuclei $^{48}$Ca, $^{50}$Ti, $^{52,54}$Cr, $^{58}$Fe, $^{59}$Co, $^{64}$Ni, $^{65}$Cu, and $^{70}$Zn with the spherical target nuclei $^{208}$Pb and $^{209}$Bi \cite{h,102jinr,102jinr1,104gsi,104gsi-a,104lbnl,106gsi,106lbnl,108gsi,108riken,110riken,110lbnl,112gsi,112riken,112riken1,folden,m}. The nuclei $^{48}$Ca, $^{50}$Ti, and $^{70}$Zn are spherical in the ground state, while the ground states of $^{52,54}$Cr, $^{58}$Fe, $^{59}$Co, $^{64}$Ni, and $^{65}$Cu are weakly-deformed \cite{msis}. It is well-known that the nucleus-nucleus interaction potential depends on the nucleon density distributions and deformations of the interacting nuclei \cite{dp,dn,dnest,dpil,dpil2,dms,ds17,dps}. The deformation of heavy nucleus effects stronger on the nucleus-nucleus potential than the deformation of the light nuclei in the case of asymmetric interacting system. Due to both the very small deformations of projectile nuclei and the weak effect of the deformation of the light nucleus on the interaction potential we may neglect by the small deformations of the projectile nuclei in the case of the cold fusion reactions. Therefore, we can consider that the spherical nuclei are participating in the cold fusion reactions.

The cross section of the SHN synthesis in collisions of spherical nuclei with the subsequent emission of $x$ neutrons from the formed compound nucleus in competition with the fission is given as
\begin{eqnarray}
\sigma^{xn}(E)=\frac{\pi \hbar^2}{2\mu E} \sum_\ell (2 \ell +1) T(E,\ell) P(E,\ell) W^{xn}(E,\ell).
\end{eqnarray}
Here $\mu$ and $E$ are the reduced mass and the collision energy of incident nuclei in the center of the mass system, respectively. $T(E,\ell)$ is the transmission coefficient through the fusion barrier formed by the Coulomb, centrifugal, and nuclear parts of the nucleus-nucleus interaction, $P(E,\ell)$ is the probability of compound nucleus formation, and $W^{xn}(E,\ell)$ is the survival probability of the compound nucleus related to the evaporation of $x$ neutrons in a competition with the fission. In a case of the cold fusion reactions $x$ equals $1$ or $2$ and rarely $3$ or $4$. The next subsections are devoted to detail description of our approaches to the calculations of $T(E,\ell)$, $P(E,\ell)$, and $W^{xn}(E,\ell)$, respectively.

\subsection{The transmission through the fusion barrier}

The total potential between spherical nuclei with the numbers of proton $Z_1$ and $Z_2$ is
\begin{eqnarray}
V_\ell(r) = \frac{Z_1 Z_2 e^2}{r} + V_{\rm N}^{\rm sph}(r) + \frac{\hbar^2 \ell(\ell+1)}{2\mu r^2}.
\end{eqnarray}
Here $r$ is the distance between centers of mass of nuclei, $e$ is the charge of proton, $V_{\rm N}^{\rm sph}(r)$ is the nuclear part of the nucleus-nucleus potential, and $\ell$ is the value of orbital angular momentum in $\hbar$ units.

The total interaction potential energy of two spherical nuclei can be approximated around the barrier by the parabola. The transmission coefficient through a parabolic barrier \cite{hw} is known exactly and given by
\begin{eqnarray}
T_{\rm par}(E,B^{\rm fus}_\ell,\ell)=1/\left[1+\exp{ \left( \frac{-2 \pi ( E-B^{\rm fus}_\ell ) }{\hbar \omega_\ell } \right) } \right],
\end{eqnarray}
where $B_\ell^{\rm fus}=B_0^{\rm fus}+\frac{\hbar^2 \ell(\ell+1)}{2\mu r_\ell^2}$ is the barrier height of the potential, $r_\ell$ is the barrier radius, and
$ \hbar \omega_\ell= \left. \left[ - \frac{\hbar^2}{\mu} \frac{d^2 V^{\rm fus}_\ell(r)}{ dr^2} \right]^{1/2} \right|_{r=r_\ell} $ is the curvature of the barrier.

The fusion barrier distribution simulated the realistic multichannel coupling is often taken into account at an evaluation of the sub-barrier heavy-ion fusion cross sections \cite{dhrs,motstef} and SHN formations \cite{zg,dns4b,dns4c,dns4d,dns5}. In this case the total transmission coefficient is given as
\begin{eqnarray}
T(E,\ell) = \int_{B_1}^{B_2} dB \; T_{\rm par}(E,B,\ell) \; f(B,B^{\rm fus}_\ell),
\end{eqnarray}
where $f(B,B^{\rm fus}_\ell) = \frac{1}{g \sqrt{\pi} } \exp{\left[ -\left(\frac{B-B^{\rm fus}_\ell}{g} \right)^2 \right]}$ is the barrier distribution function, which usually approximates by the Gauss function \cite{zg,dns4b,dns4c,dns4d}. The typical values of the barrier distribution width $g$ are several MeVs \cite{zg,dns4b,dns4c,dns4d}.

In the case of sub-barrier energies $E \ll B^{\rm fus}_\ell$ Eq. (3) can be approximated in the form $T_{\rm par}(E,B^{\rm fus}_\ell,\ell) \approx \exp{ \left( \frac{2 \pi ( E-B^{\rm fus}_\ell ) }{\hbar \omega_\ell } \right) }.$
Substituting this expression into Eq. (4) and extend the limits of the integral to infinity we get $T(E,\ell) \approx \exp{ \left( \frac{2 \pi ( E-(B^{\rm fus}_\ell - \Delta_B) ) }{\hbar \omega_\ell } \right) },$ where $\Delta_B = \pi g^2/(2\hbar \omega_\ell)$ is the the shift of the barrier value due to the barrier distribution. Using this property, we approximate the transmission coefficient through the distribution of the parabolic barriers for any values of $E$ as
\begin{eqnarray}
T(E,\ell) \approx 1/\left[1+\exp{ \left( \frac{-2 \pi ( E-(B^{\rm fus}_\ell - \Delta_B)) }{\hbar \omega_\ell } \right) } \right].
\end{eqnarray}
It is obvious that the values of $T(E,\ell)$ obtained using this formula at energies far from the barrier value are the same as the calculated with the help of Eqs. (3)-(4). The values of $T(E,\ell)$ calculated by the approximate formula (5) deviate from the exact values using Eqs. (3)-(4) at energies around the barrier. This deviation decreases with a decreasing $g$. The application of the parameter $\Delta_B$ is simpler than a numerical integration of the barrier distribution (4) and leads to the same effect.

We should know the nucleus-nucleus potential for evaluating the transmission coefficient using Eq. (5). The nuclear part of the nucleus-nucleus potential consists of the macroscopic and the shell-correction contributions \cite{d2015}
\begin{eqnarray}
V_{\rm N}^{\rm sph}(r) = V_{\rm macro}(r) + V_{\rm sh}(r) .
\end{eqnarray}

The macroscopic part $V_{\rm macro}(r)$ of the nuclear interaction of nuclei is related to the macroscopic density distribution and the nucleon-nucleon interactions of colliding nuclei. It is the Woods-Saxon form at $r>R_{\rm t}$ \cite{d2015}
\begin{eqnarray}
V_{\rm macro}(r) = \frac{v_1 C+ v_2 C^{1/2}}{1+\exp[(r-R_{\rm t})/(d_1 + d_2/C)]} .
\end{eqnarray}
Here $v_1=-27.190$ MeV fm$^{-1}$, $v_2=-0.93009$ MeV fm$^{-1/2}$, $d_1=0.78122$ fm, $d_2= - 0.20535$ fm$^2$, $C=R_1 R_2/R_{\rm t}$ is in fm, $R_{\rm t}=R_1+R_2$, $R_i=1.2536 A_i^{1/3}-0.80012 A_i^{-1/3} -0.0021444/A_i$ is the radius of $i$-th nucleus in fm, $i=1,2$, and $A_i$ is the number of nucleons in nucleus $i$.

The shell-correction contribution $V_{\rm sh}(r)$ to the potential is related to the shell structure of nuclei, which is disturbed by the nucleon-nucleon interactions of colliding nuclei. When the nuclei approach each other, the energies of the single-particle nucleon levels of each nucleus are shifted and split due to the interaction of nucleons belonging to different nuclei. This changes the shell structures of both nuclei at small distances between them. Therefore, the shell-correction contribution into the total nuclear interaction of nuclei is introduced in Ref. \cite{d2015}. This representation of the total nuclear potential energy of two nuclei is similar to the Strutinsky shell-correction prescription for the nuclear binding energy \cite{s,s1,s2,s3}. The shell-correction part of the potential at $r>R_{\rm t}$ is given as \cite{d2015}
\begin{eqnarray}
V_{\rm sh}(r) = [\delta E_1 + \delta E_2 ] \left[\frac{1}{1 + \exp{ \left( \frac{R_{\rm sh}-R}{d_{\rm sh}} \right)}}-1 \right],
\end{eqnarray}
where $R_{\rm sh}= R_{\rm t} - 0.26$ fm, $d_{\rm sh} = 0.233$ fm, and
\begin{eqnarray}
\delta E_i = B^{\rm exp}_i - B^{\rm m}_i
\end{eqnarray}
is the phenomenological shell correction for nucleus $i$.
\begin{eqnarray}
B^{\rm m}_i &=& -15.86864 A_i+21.18164 A^{2/3}_i-6.49923 A^{1/3}_i \nonumber \\
 &+&\left[\frac{N_i-Z_i}{A_i}\right]^2 \left[26.37269 A_i -23.80118 A^{2/3}_i \right. \nonumber \\ &-& \left. 8.62322 A^{1/3}_i \right] \\
 &+& \frac{Z^2_i}{A^{1/3}_i} \left[ 0.78068- 0.63678 A^{-1/3}_i \right] + P_p+P_n \nonumber
\end{eqnarray}
is the macroscopical value of the binding energy in MeV founded in phenomenological approach, $B_{\rm exp}$ is the binding energy of nucleus in MeV obtained using the evaluated atomic masses \cite{be}, $P_{p(n)}$ are the proton (neutron) pairing terms, which equal to $P_{p(n)}=5.62922 (4.99342) A^{-1/3}_i$ in the case of odd $Z$ ($N$) and $P_{p(n)}=0$ in the case of even $Z_i$ ($N_i$), $N_i$ is the number of neutrons in nucleus $i$.

The parametrization of $V_{\rm N}^{\rm sph}(r_\ell)$ from Ref. \cite{d2015} is used in our model, because the barrier heights $B^{\rm fus}_0$ calculated with the help of this parametrization agree well with the empirical values of barrier heights for light, medium and heavy nucleus-nucleus systems \cite{d2015,bass,empbar1,empbar2a}. The values of the barriers for the spherical systems $^{48}$Ca, $^{48,50}$Ti, $^{52}$Cr, $^{54}$Cr, $^{56,58}$Fe, $^{64}$Ni, $^{70}$Zn + $^{208}$Pb leading to the SHN obtained in our approach are presented in Table 1. These values of the barriers well agree with the available values of the barriers derived from an analysis of the experimental data for the quasi-elastic backscattering \cite{mitsuoka}, see Table 1. Note, this parametrization is also successfully used for the description of the fragment mass distribution at the fission of highly-excited nuclei \cite{dms,ds17} and the ternary fission \cite{dps}. So, the values barrier heights obtained in our approach are reliable.

Using Eqs. (2), (5)-(10) we can evaluate the capture cross section
\begin{eqnarray}
\sigma_{\rm cap}(E)=\frac{\pi \hbar^2}{2\mu E} \sum_\ell (2l+1) T(E,\ell) .
\end{eqnarray}
The capture cross section is related to the fusion barrier penetration. The capture cross section coincides with the compound nucleus production cross section in the case of collisions of light and medium nuclei, when the decay of DNS on fragments and the quasi-fission process give negligible contributions \cite{dhrs,motstef,den-subfus,ds2019,empbar2a}. In the case of collisions of heavy nuclei, the capture cross section links to the formation of the DNS.

\begin{table}
\caption {
\label{tab1}The values of barrier heights between spherical nuclei $B^{\rm fus}_\ell$ for $\ell=0$, Q values of the compound nucleus formation $Q_{\rm CN}$ obtained using the evaluated atomic masses \cite{be}, the excitation energies of the compound nucleus at the collision energies equal to the barrier heights $E_{\rm bar}^*=E+Q_{\rm CN}$, and the available values of the barrier heights $B_{\rm qebs}$ derived from an analysis of the experimental data for the quasi-elastic backscattering \cite{mitsuoka}. All values are given in MeV.}
\begin{center}
\begin{tabular}{|c|cccc|}
\hline
Collision systems & $B^{\rm fus}_0$ & $-Q_{\rm CN}$ & $E_{\rm bar}^*$ & $B_{\rm qebs}$ \\
\hline
$^{48}$Ca + $^{208}$Pb & 172.5 & 153.8 & 18.7 & \\
$^{48}$Ti + $^{208}$Pb & 191.0 & 164.5 & 26.5 & 190.1 \\
$^{50}$Ti + $^{208}$Pb & 189.8 & 169.5 & 20.3 & \\
$^{52}$Cr + $^{208}$Pb & 207.2 & 183.7 & 23.5 & \\
$^{54}$Cr + $^{208}$Pb & 206.0 & 187.1 & 19.0 & 205.8 \\
$^{56}$Fe + $^{208}$Pb & 223.2 & 201.9 & 21.3 & 223.0 \\
$^{58}$Fe + $^{208}$Pb & 222.0 & 205.0 & 17.0 & \\
$^{64}$Ni + $^{208}$Pb & 236.6 & 224.9 & 11.9 & 236.0 \\
$^{70}$Zn + $^{208}$Pb & 251.1 & 244.2 & 6.9 & 250.6 \\
\hline
\end{tabular}
\end{center}
\end{table}

\subsection{The probability of compound nucleus formation}

\subsubsection{Expression for the probability of compound nucleus formation}

The height of the fusion barrier between incident spherical nuclei is high. The collision energy of the two nuclei starts to dissipate just before the barrier. After passing the incident fusion barrier the nuclei locate in the capture well, which is close to the contact distance of nuclei. The nuclei form the DNS in the capture well.
The DNS formed in capture well is the injection point for subsequent stages of the SHN formation and the DNS evolution.

The kinetic energy related to the relative motion of the colliding nuclei is quickly dissipated at the initial collision stage, when the tails of nucleon densities of the nuclei is starting to overlap. As a result, the kinetic energy of the relative motion of nuclei transfers into the intrinsic energy of the DNS \cite{v2,fl}. Therefore, the subsequent stages of the SHN synthesis can be considered in the framework of the statistical approach. In this approach the probability of compound nucleus formation is linked to the ratio of the decay widths of different processes. The widths considered in this subsection are defined in the framework of the Bohr-Wheeler transition state statistical approach \cite{bw}.

The DNS formed in capture well may decay into different channels as, for example, the spherical or deformed incident nuclei, new DNSs formed at the transfer of nucleons between the incident nuclei, and formation of the compound nucleus. Due to the transfer of nucleons between nuclei the DNS may decay into the more symmetric nucleus-nucleus systems with subsequent decay on the deformed nuclei or the more asymmetric nucleus-nucleus systems with subsequent formation of the compound nucleus. The DNS may decay into the compound nucleus by the smooth shape evolution too. The corresponding decay branches of the DNS are linked to the respective decay widths and barriers.

The probability of specific decay process is related to the passing through the barrier in competition to other processes. The compound nucleus is formed by passing the barrier of the fusion way and the DNS barrier related to the transfer of nucleons from the light nucleus to the heavy one. The passing through other barriers are linked to the decay of the DNS on scattered nuclei.

Therefore, the probability of the compound nucleus formation from the DNS is determined in our model as the ratio of the widths leading to the compound nucleus to the total decay widths of the DNS, i.e.
\begin{eqnarray}
P(E,\ell) = \frac{\Gamma_{\rm CN}^{\rm DNS, f}(E,\ell) + \Gamma_{\rm CN}^{\rm DNS, tr}(E,\ell)}{\Gamma^{\rm tot}_{\rm CN}(E,\ell)} .
\end{eqnarray}
Here
\begin{eqnarray}
\Gamma^{\rm tot}_{\rm CN}(E,\ell)&=&\Gamma_{\rm CN}^{\rm DNS, f}(E,\ell)+\Gamma_{\rm CN}^{\rm DNS, tr}(E,\ell) + \Gamma_{\rm DIC}^{\rm DNS}(E,\ell) \nonumber \\ &+& \Gamma_{\rm sph}^{\rm DNS}(E,\ell) +\Gamma_{\rm def}^{\rm DNS}(E,\ell)
\end{eqnarray}
is the total decay widths of the DNS. $\Gamma_{\rm CN}^{\rm DNS, f}(E,\ell)$ is the width related to the compound nucleus formation through the fusion path by the smooth evolution of shape of nuclear system. $\Gamma_{\rm CN}^{\rm DNS, tr}(E,\ell)$ is the width of the compound nucleus production through the multi-nucleon transfer from the light to heavy nuclei. This is the DNS path of compound nucleus formation used in the various versions of the DNS model \cite{dns1,dns1c,dns1aa,dns1aa1,dns1aa2,dns1aa4,dns2a,adamian,dns3,dns31,dns32,dns33,dns3a,dns4b,dns4c,dns4d,dns5}. $\Gamma_{\rm DIC}^{\rm DNS}(E,\ell)$ is the width of the DNS decay into two nuclei with nucleon transfer from heavy to light nuclei. During this process the DNS decays into scattered nuclei, which are close to the incident nuclei. This process is the deep-inelastic collisions (DIC). $\Gamma_{\rm sph}^{\rm DNS}(E,\ell)$ and $\Gamma_{\rm def}^{\rm DNS}(E,\ell)$ are the width of the DNS decay into incident spherical or deformed nuclei, respectively. The widths $\Gamma_{\rm sph}^{\rm DNS}(E,\ell)$ and $\Gamma_{\rm def}^{\rm DNS}(E,\ell)$ are connected to the different quasi-elastic decay modes of the DNS.

Eq. (12) is written for the case of a direct coupling of the DNS to the equilibrium shape of compound nucleus. In this case the potential energy landscape has the valley, the bottom of which is directly connecting the system of contacting nuclei and the equilibrium shape of compound nucleus, see, for example, the landscapes presented in Fig. 1. If the final point of the valley started from the DNS is related to the non-equilibrium shape of compound nucleus and the equilibrium shape of compound nucleus locates in another valley, see, for example, the landscapes presented in Fig. 2, then the intermediate state should be introduced. The intermediate state can decay into both the compound nucleus in equilibrium shape and to the quasi-fission fragments. Such case of the compound nucleus formation will be considered in Sec. 2.B.3.

According to Eqs. (12)-(13), the compound nucleus in our model can be formed by both the fusion and DNS ways. If we put $\Gamma_{\rm CN}^{\rm DNS, f}(E,\ell)=0$ and introduce $\Gamma_{\rm quasi-fission}(E,\ell)=\Gamma_{\rm DIC}^{\rm DNS}(E,\ell)+\Gamma_{\rm sph}^{\rm DNS}(E,\ell)+\Gamma_{\rm def}^{\rm DNS}(E,\ell)$ then our probability of compound nucleus formation (12)-(13) equals to the one in Refs. \cite{v,dns1,dns1c,adamian}. Here we use the original name of the width $\Gamma_{\rm quasi-fission}(E,\ell)$ used in Refs. \cite{v,dns1,dns1c,adamian}. Note, in our model the quasi-fission process is related to the decay of one-body nuclear shape on the fission fragments bypassing the formation of the compound nucleus with the equilibrium shape. (Recently, the authors of the DNS model have used the master equation approach for evaluation of the probability of compound nucleus formation \cite{dns2a,dns3,dns31,dns32,dns33,dns3a,dns4b,dns4c,dns4d,dns5}. However, the old \cite{v,dns1,dns1c,adamian} and new \cite{dns2a,dns3,dns31,dns32,dns33,dns3a,dns4b,dns4c,dns4d,dns5} approaches of the DNS model have the same mechanism of the compound nucleus formation. In our model the probability of compound nucleus formation is described by the ratio of the decay widths in Eqs. (12)-(13). This is very convenient, because all decay processes are considered in the same approach.)

The shape and properties of the potential energy landscape related to the fusion trajectory of compound nucleus formation is discussed in next subsection.

\subsubsection{The landscape of potential energy related to the fusion path from the DNS to compound nucleus}

The DNS is formed in the collision of two spherical nuclei in the case of cold fusion reactions. The DNS after formation can evolve to the compound nucleus or divide on two spherical or deformed nuclei. The division of the DNS on two deformed fragments can be linked as to the quasi-fission as to the immediately decay of the DNS. Therefore, the shapes of nucleus related to the analysis of the trajectory from the DNS to the compound nucleus with possibility of the quasi-fission should include the two spherical or deformed nuclei as well as the spherical, well-deformed, and pre-ruptured one-body nuclear shapes. The parametrization describing such different shapes should be simple as possible. The axial-symmetric parametrization
\begin{eqnarray}
\rho= \left\{ \begin{array}{ll}
0, & \textrm{if} \; z<-a,\\
b \sqrt{1-(z/a)^2}, & \textrm{if} \; -a \leq z \leq z_0, \\
c \sqrt{1-((z-R)/d)^2}, & \textrm{if} \; z_0 \leq z \leq R+d,\\
0, & \textrm{if} \; z>R+d
\end{array} \right.
\end{eqnarray}
is satisfied the proposed conditions. Here $\rho$ and $z$ are the cylindrical coordinates. The parametrization depends on the 6 parameters $a,b,c,d,z_0$, and $R$.

The radius $\rho$ should be continued at the point $z_0$, therefore
\begin{eqnarray}
b \sqrt{1-(z_0/a)^2} = c \sqrt{1-((z_0-R)/d)^2}.
\end{eqnarray}
This equation couples the two parameters of the parametrization, for example, $z_0$ and $R$. The total volume of the nuclear system should be conserved during the shape evolution of the nuclear system. As a result of these constraints, the shape parametrization has 4 independent parameters.

We fix values $a = b = R_1$, where $R_1$ is the radius of the light incident nucleus. Due to this fixing the shape parametrization (14) depends only on two independent parameters. Note that two independent parameters are often used for description of the compound nucleus formation at evolution of the one-body form, see, for examples, Refs. \cite{shen1,boilley,torres2,torres,torres1}.

The two touching nuclei is described by Eq. (14) at $z_0=R_1$. During fusion the nuclei get close and the value of $z_0$ is smoothly reduced. The radius of the neck connecting nuclei is rising from 0 at $z_0=R_1$ up to the radius of the light nucleus $R_1$ at $z_0=0$. After that the heavy nucleus absorbs the light one and $z_0$ approaches to $-R_1$. The light nucleus is fully absorbed by the heavy one at $z_0=-R_1$.

It is useful to consider two independent variables $z_0$ and $\beta_2$ for specification of the shape of fusing nuclei. Parameter $\beta_2$ is coupled to the ratio $d/c$ by the equation $d/c=[1+\beta_2 Y_{20}(\theta=0^\circ)]/[1+\beta_2 Y_{20}(\theta=90^\circ)]$, where $Y_{20}(\theta)$ is the spherical harmonic function \cite{vmk}. Parameter $\beta_2$ describes the quadrupole deformation of heavy nucleus in the case of the touching nuclei at $z_0=R_1$ or the quadrupole deformation of the one-body shape at $z_0=-R_1$. (Here we neglect the difference between the equal-volume shapes of the axial-symmetric ellipsoid and the nucleus with the surface radius $R(\theta)=R_0[1+\beta_2 Y_{20}(\theta)]$. Such shapes are very close to each other at small deformations.) Our shape parametrization at $z_0=-R_1$ describes the fission process related to the evolution of the quadrupole deformation too. Note that the quadrupole deformation is successfully used for a discussion of the fission process by Bohr and Wheeler \cite{bw}.

Our two-parameters parametrization is useful for the simultaneous description of various nuclear shapes related to the fusion of asymmetric nuclei, compound nucleus formation, fission and quasi-fission. This parametrization has not used anywhere, see, for example, Ref. \cite{hasse} devoted to the compilation of various nuclear shapes used in literature.

For every value $z_0$ and $\beta_2$ we find parameters $R^{fit}$ and $\beta_L^{fit}$, at which the parametrization
\begin{eqnarray}
R(\theta)=R^{fit}[1+\sum_{L=2}^9 \beta_L^{fit} Y_{L0}(\theta)]
\end{eqnarray}
fits the shape described by Eq. (14). Using obtained values of parameters $R^{fit}$ and $\beta_L^{fit}$ we calculate the shell correction energies as the function of the parameters $z_0$ and $\beta_2$ by using the code WSBETA \cite{wsbeta}. This code uses a Woods-Saxon potential with a 'universal' parameter set and the parametrization of the nuclear shape in the form $R(\theta) \propto [1+\sum_{L=2}^9 \beta_L^{fit} Y_{L0}(\theta)]$. The radius parameter of the Woods-Saxon potential is fixed by the 'universal' parameter set \cite{wsbeta}. Therefore, the parameters $z_0$ and $\beta_2$ are coupled to the deformation parameters $\beta_L^{fit}$ only. The residual pairing interaction is calculated by means of the Lipkin-Nogami method \cite{paring}. The macroscopic part of the deformation energy is evaluated using the Yukawa-plus-exponential potential \cite{macro}.

The dependencies of the potential energy landscape on the variables $z_0$ and $\beta_2$ for the cold-fusion systems $^{50}$Ti, $^{52,54}$Cr, $^{58}$Fe, $^{64}$Ni, $^{70}$Zn, and $^{78}$Ge + $^{208}$Pb are presented in Figs. 1--2. The dependencies of the potential energy of nuclei on $\beta_2$ at $z_0=-R_1$ presented in these Figs. are typical for the fissioning nucleus. So, we see the ground-state well and the fission barrier along the line $z_0=-R_1$ in these Figures.

There are many heavy and super-heavy nuclei with the two-hampered fission barrier \cite{bs,bl,sob,sob1,etfsi,d-uhe,mol,snp,pnsk,pnsk-a,kowal,kowal-a,kjs,afanas,afanas-a,agbemava,warga}. The values of $\beta_2$ at the ground state of the compound nucleus, which we can see in Figs. 1--2 at $z_0 \approx -R_1$, are close to the corresponding ones obtained in Ref. \cite{kjs}. The values of the inner fission barrier evaluated from Figs. 1-2 at $z_0 \approx -R_1$ are close to the ones obtained in Refs. \cite{kowal,kowal-a,kjs,etfsi,mol} in the framework of the shell-correction approach. Note that Refs. \cite{sob,sob1,mol,kowal,kowal-a,kjs} are devoted to the accurate calculations of the ground state and saddle point properties of SHN using a rich set of the multipole deformations. Our deformation space is limited by two independent variables, therefore the agreement of the fission barrier values extracted from Figs. 1-2 with the results obtained in the framework of other models is approximate.

\begin{figure}
\includegraphics[width=6.8cm]{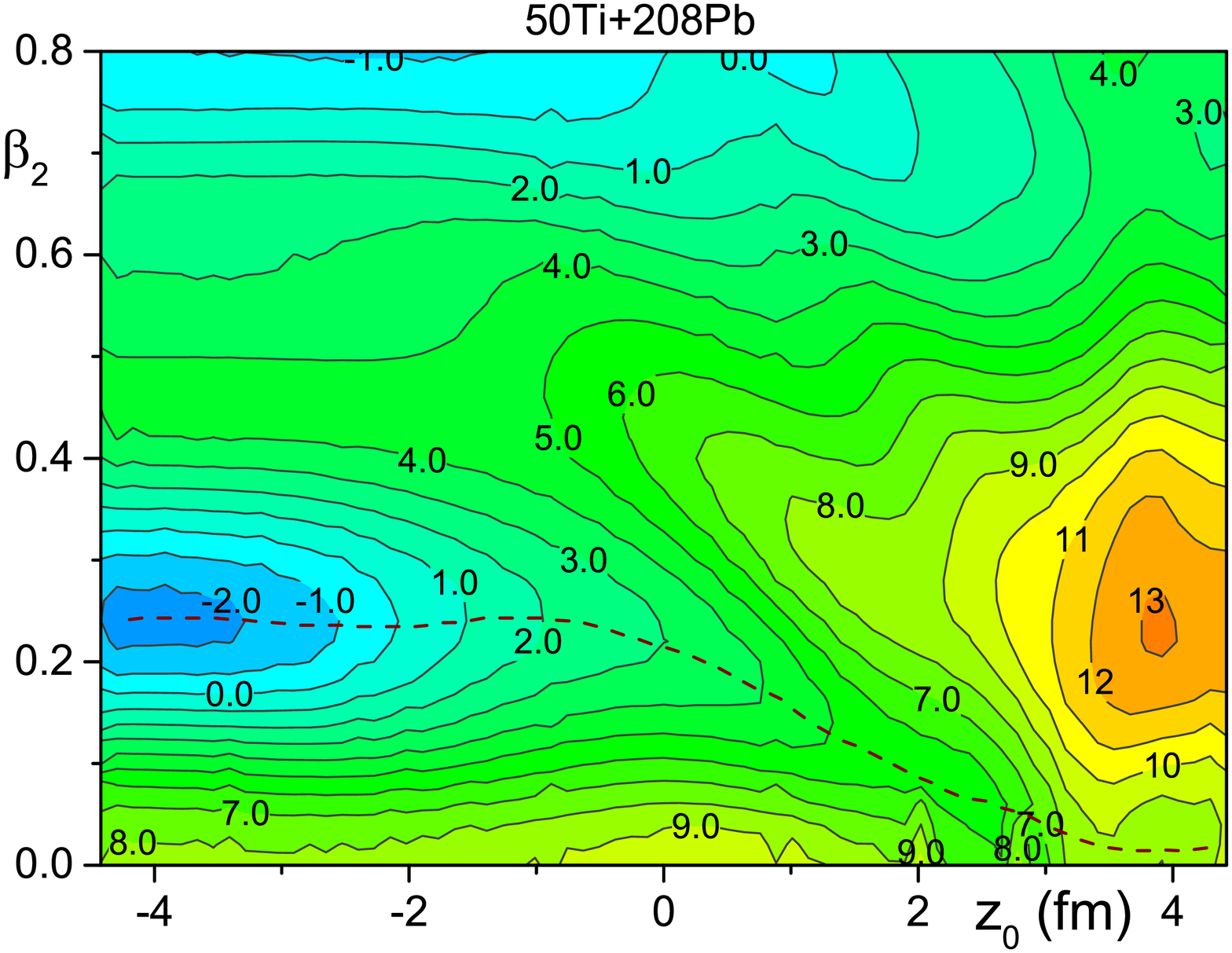}
\includegraphics[width=6.8cm]{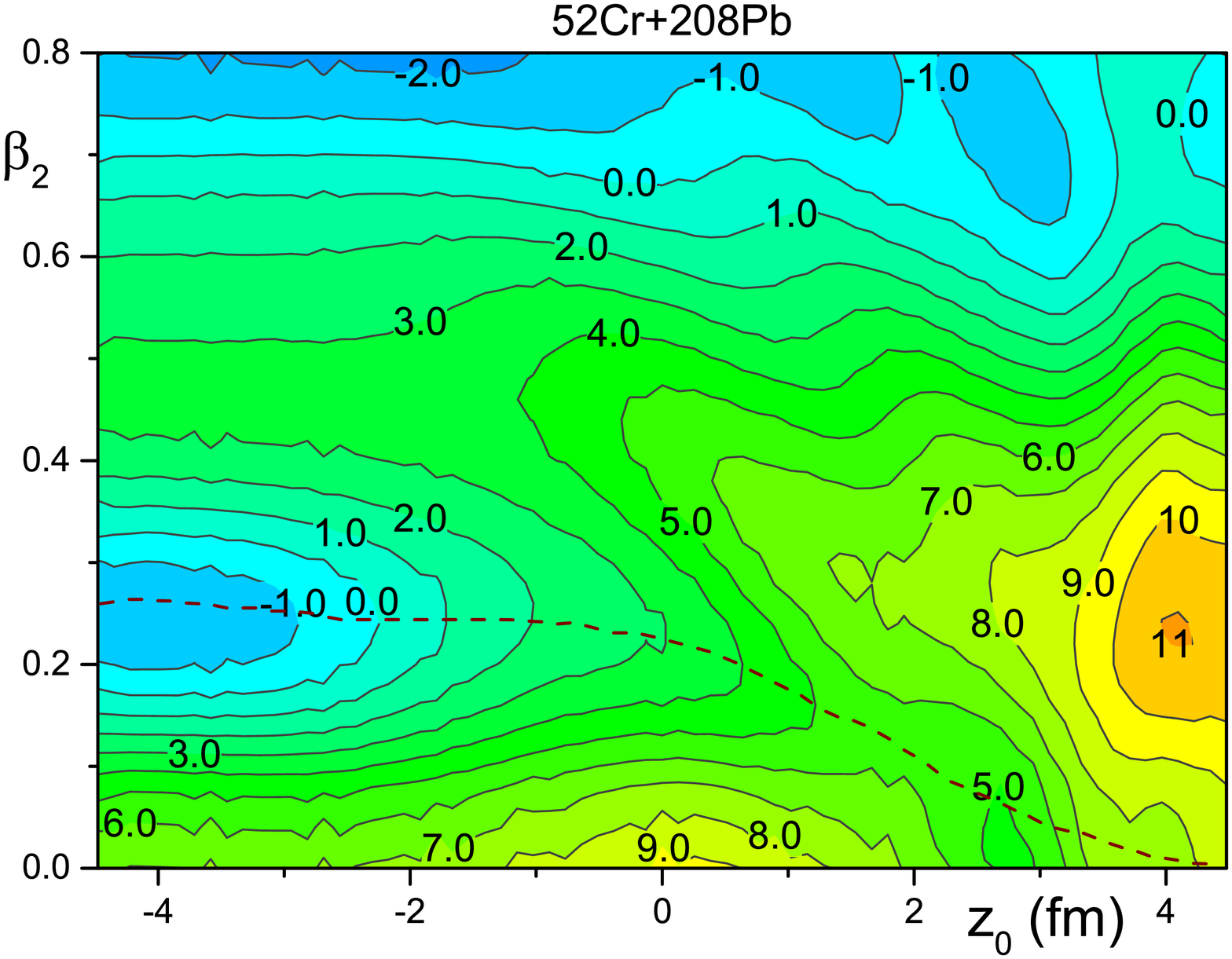}
\includegraphics[width=6.8cm]{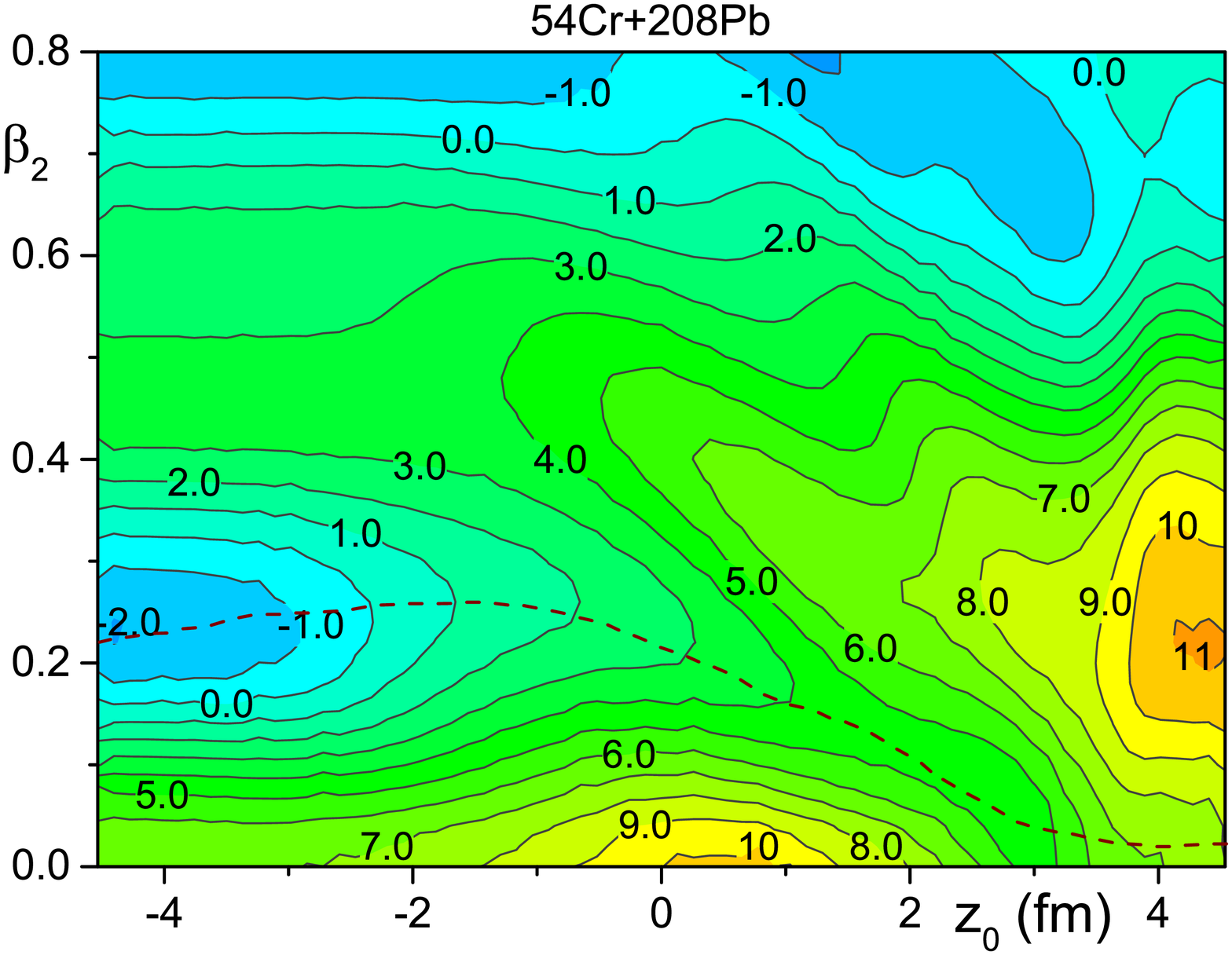}
\includegraphics[width=6.8cm]{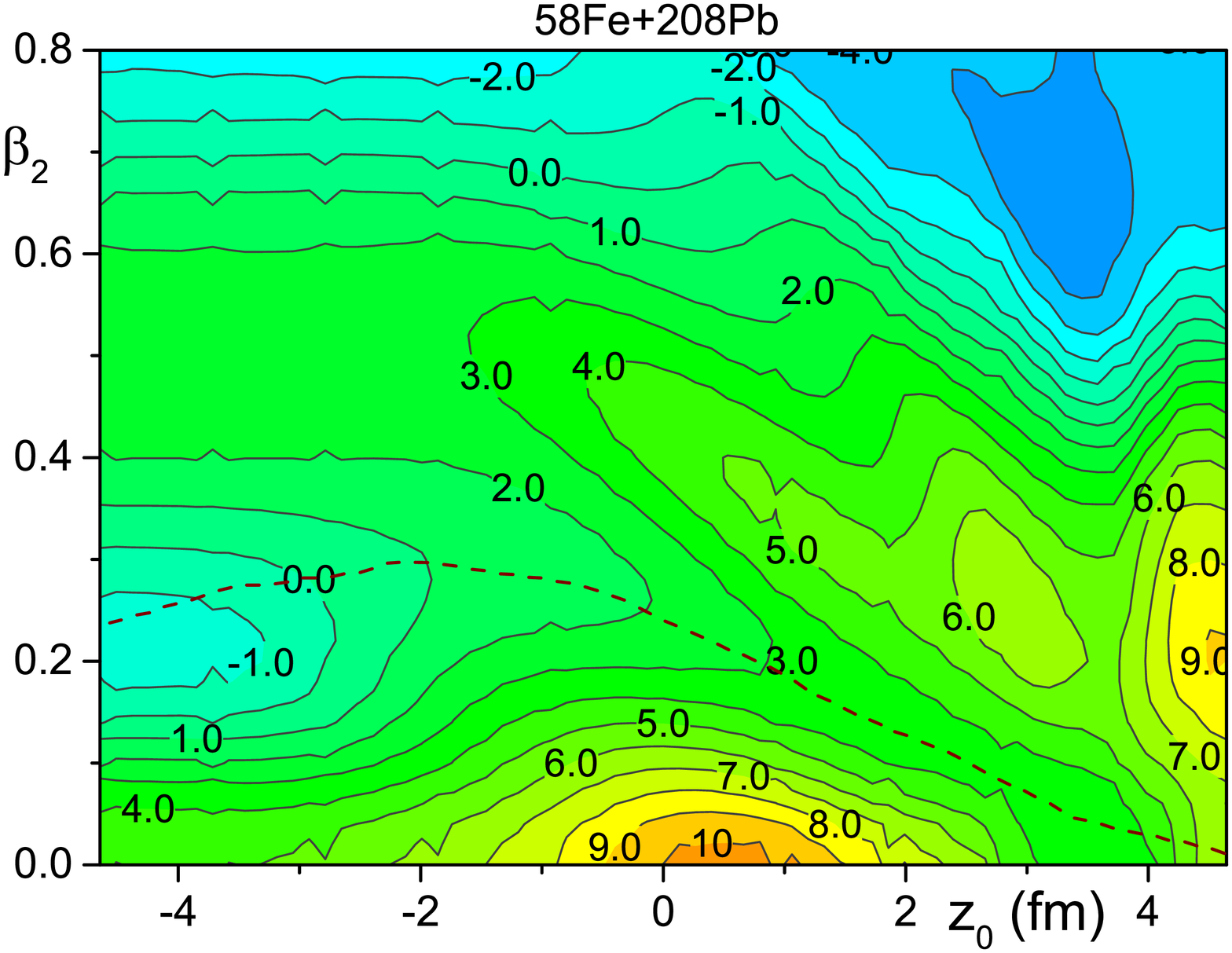}
\caption{\label{fig1} The potential energy landscape in depending on the variables $z_0$ and $\beta_2$ for cold-fusion systems $^{50}$Ti, $^{52,54}$Cr, and $^{58}$Fe + $^{208}$Pb. The dashed lines are the trajectories of the compound nucleus formation, which are drawn by eye. }
\end{figure}

\begin{figure}
\includegraphics[width=6.8cm]{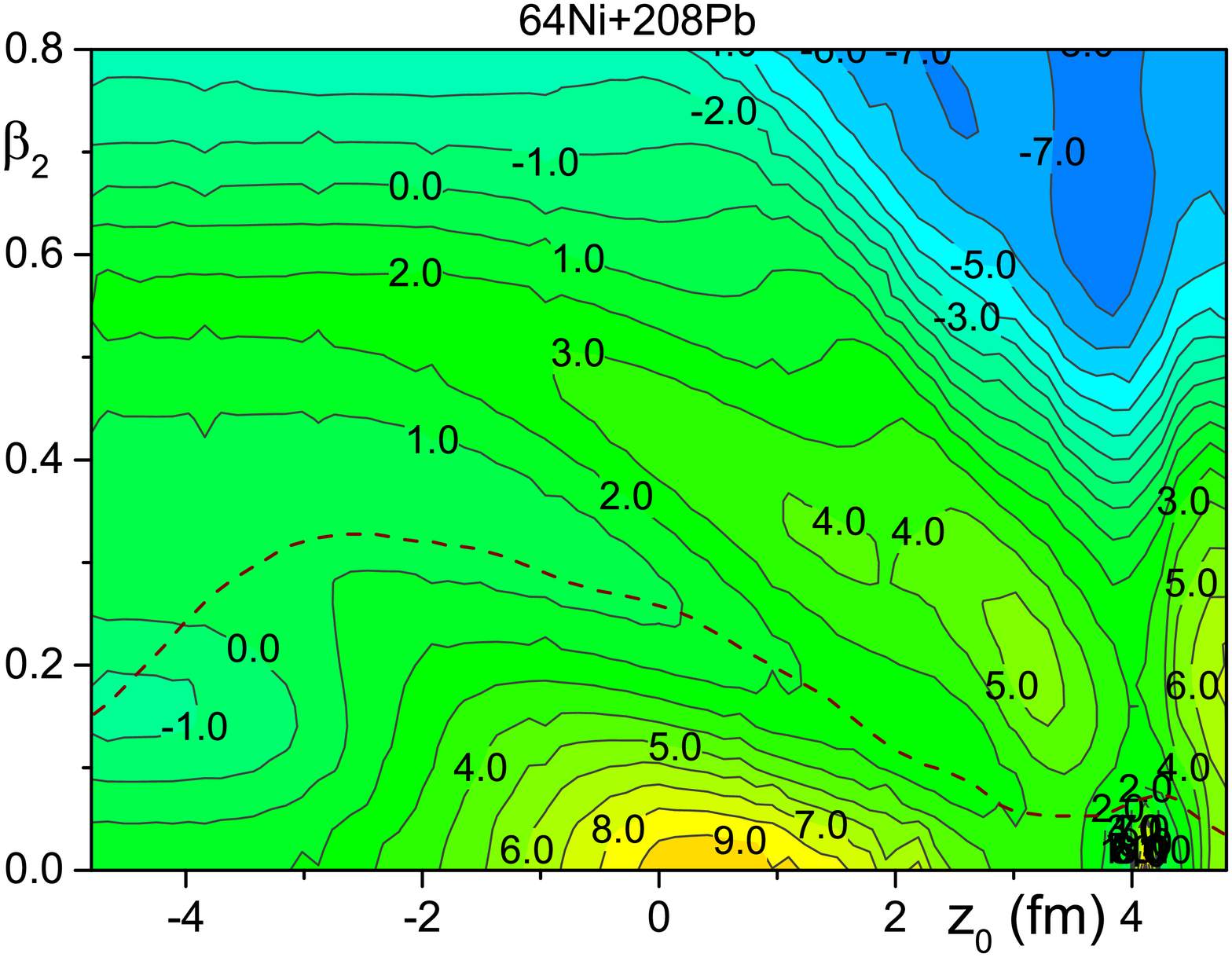}
\includegraphics[width=6.8cm]{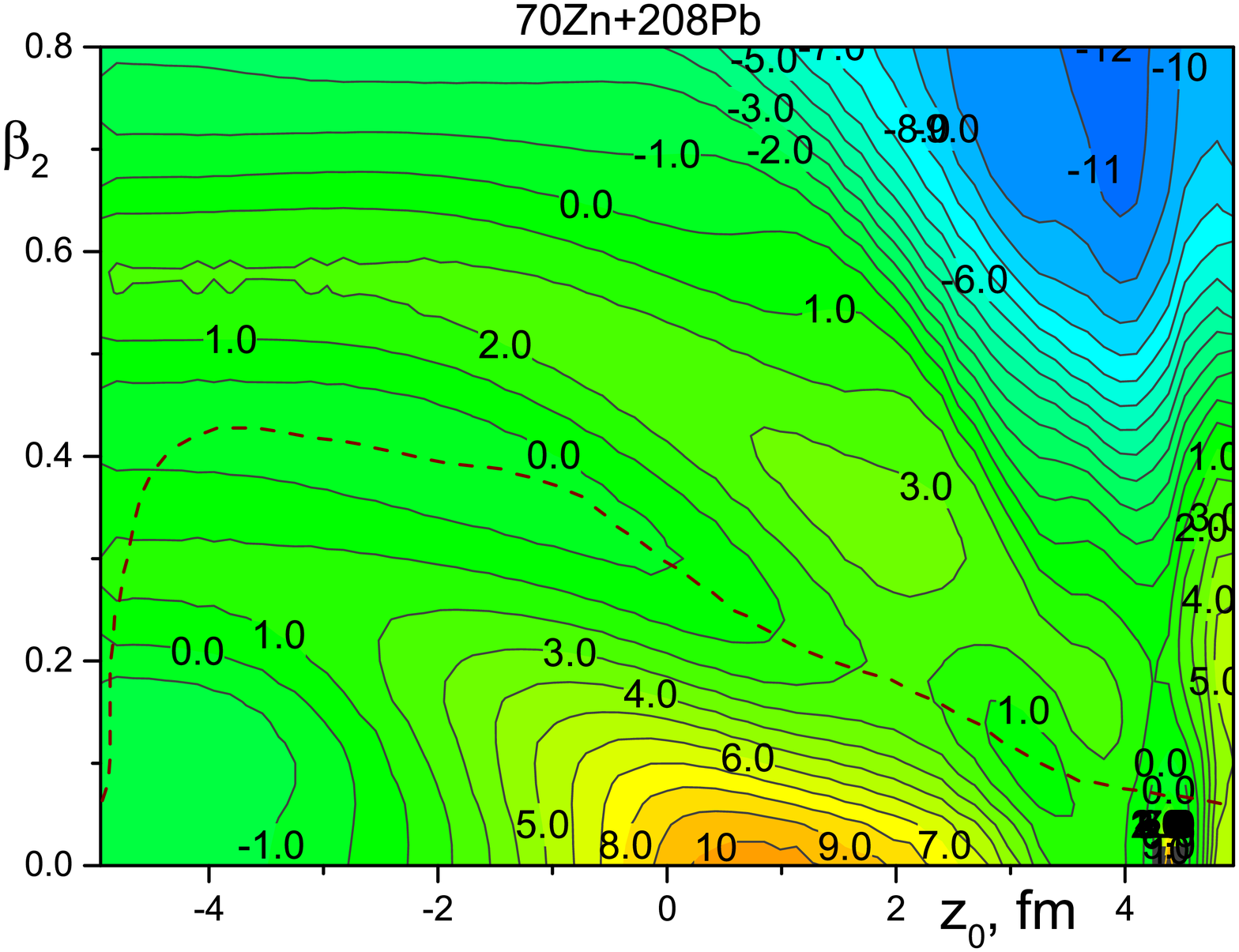}
\includegraphics[width=6.8cm]{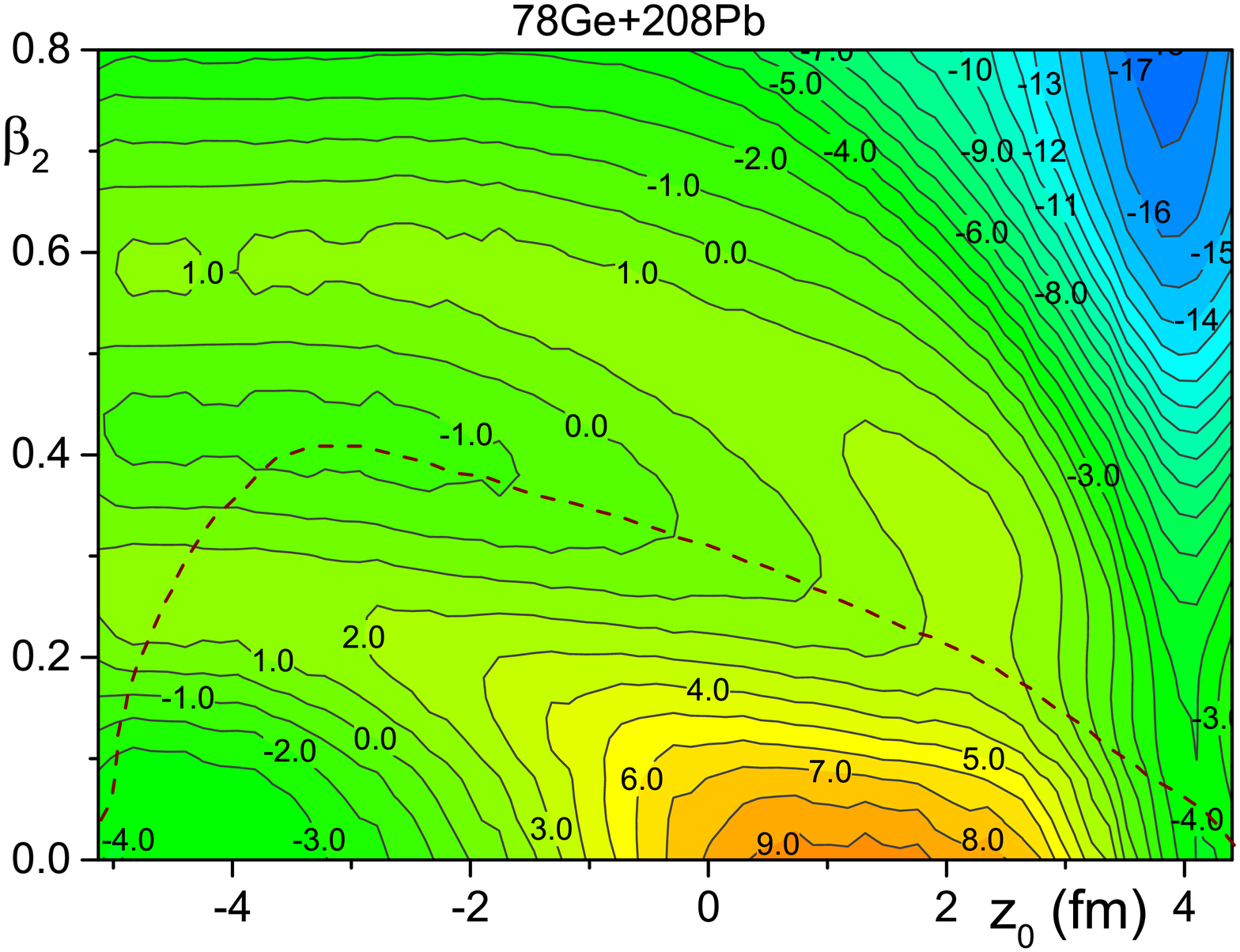}
\caption{\label{fig2} The potential energy landscape in depending on the variables $z_0$ and $\beta_2$ for cold-fusion systems $^{64}$Ni, $^{70}$Zn, and $^{78}$Zn + $^{208}$Pb. The dashed lines are the compound nucleus formation trajectories, which are drawn by eye.}
\end{figure}

The dependence of the potential energy of two contacting nuclei on the value of quadrupole deformation of heavy nucleus is presented in Figs. 2--3 at $z_0=R_1$. We see the potential energy has a local minimum at small values of $\beta_2 \approx 0$ at $z_0=R_1$, which is related to the spherical ground-state shape of $^{208}$Pb. Note that the similarity of the shapes described by the parametrizations (14) and (16) is worsened at $z_0 \rightarrow R_1$. Nevertheless, the shape described by Eq. (16) is close to the shape of two contacting nuclei.

The landscape of potential energy is strongly changed near $z_0=R_1$ for the system $^{78}$Ge + $^{208}$Pb. As a result, the contour lines are difficultly separated, therefore we present the potential landscape for this system for $z_0\leq4.4$ fm, see Fig. 2.

The trajectory of the compound nucleus formation, which is drawn by eye in Figs. 1--2, connects the point of two contacting spherical nuclei at $z_0=R_1, \beta_2 = 0$ and the point of the ground state of the compound nucleus at $z_0 \approx -R_1$ and $\beta_2$ from the range $0 \leq \beta_2 \leq 0.25$. These trajectories for the cold-fusion systems $^{50}$Ti, $^{52,54}$Cr, $^{58}$Fe + $^{208}$Pb are located in the bottom of the valley leading to the compound nucleus formation, see Fig. 1. Similar valleys are also obtained for the cluster emission from heavy nuclei with the daughter nucleus near $^{208}$Pb in Ref. \cite{warda18}. The fusion path leading to the compound nucleus formation is also studied in Ref. \cite{moller,moller-a}.
The dependencies of the potential energies on the elongation along the fusion paths presented in Refs. \cite{moller,moller-a} for reactions $^{50}$Ti and $^{70}$Zn + $^{208}$Pb look like the ones in Figs. 1 and 2 for the these systems. Unfortunately, the direct comparison of the potentials along the fusion trajectories obtained in these different approaches is not possible due to using various shape parametrizations.

The high ridge separates the fusion valley and the quasi-fission area in Fig. 2. This ridge merges smoothly to the inner fission barrier at $z_0=-R_1$. The true quasi-fission (or fast fission) process in heavy-ion reactions is related to the fission of nuclear system before reaching the ground-state shape of the compound nucleus, while the true fission is started from the ground-state shape of the compound nucleus. The large difference between the energies at the bottom of the fusion valley and at the ridge leads to a statistical suppression of the quasi-fission process in comparison to the compound nucleus formation for systems $^{50}$Ti, $^{52,54}$Cr, $^{58}$Fe + $^{208}$Pb. Therefore, the probabilities of compound nucleus formation are well determined by Eqs. (12)-(13) for these systems.

We see the saddle points at the point $z_0 \approx 2-4$ fm and $\beta_2 \approx 0.1-0.25$ on the fusion path for systems $^{64}$Ni, $^{70}$Zn, and $^{78}$Ge + $^{208}$Pb in Fig. 2. The heights of these saddle points define the barriers for the transition from the DNS to the compound nucleus along the fusion trajectories for these systems. In contrast to this, such saddle points absent for the fusion paths for reactions $^{50}$Ti, $^{52,54}$Cr, and $^{58}$Fe + $^{208}$Pb in Fig. 1. Therefore, the barriers related to the formation of the compound nucleus along the fusion trajectory for reactions $^{50}$Ti, $^{52,54}$Cr, and $^{58}$Fe + $^{208}$Pb are defined as the highest values of the potential energies of spherical or near-spherical nuclei at the contact point, see Fig. 1. This is because the excitation energies of these systems in the contact point should be above or equal these potential energies for the successful formation of the compound nuclei.

The trajectories of the compound nucleus formation for reactions $^{64}$Ni, $^{70}$Zn, and $^{78}$Ge + $^{208}$Pb have the saddle points near the point $z_0 \approx 4$ fm and $\beta_2 \approx 0.1-0.2$, which are linked to the decay of the slightly overlapped nuclei (or the DNS) on two fragments, see Fig. 2. The decay of DNS on two fragments is also described by the widths $\Gamma_{\rm DIC}^{\rm DNS}(E,\ell)$, $\Gamma_{\rm sph}^{\rm DNS}(E,\ell)$ and $\Gamma_{\rm def}^{\rm DNS}(E,\ell)$. The width $\Gamma_{\rm def}^{\rm DNS}(E,\ell)$ is connected to the lowest value of the barrier, which is taken place for the two-body systems. The height of the saddle point near the point $z_0 \approx 4$ fm and $\beta_2 \approx 0.1-0.2$ is higher the height of the barrier related to the width $\Gamma_{\rm def}^{\rm DNS}(E,\ell)$. Therefore, we may neglect by the influence of this saddle point on both the formation of the compound nucleus and the decay of DNS on fragments.

The ridge, which separates the compound nucleus formation valley and the quasi-fission valley, merges to the outer fission barrier near $z_0 \approx -R_1$ for the cold-fusion systems $^{70}$Zn and $^{78}$Ge + $^{208}$Pb, see Fig. 2. The compound nucleus formation valley is merged with the potential energy well between the inner and outer fission barrier near $z_0 \approx -R_1$. Therefore, the intermediate state is formed in this well. The compound nucleus is formed at the decay of the intermediate state through the inner fission barrier. The quasi-fission fragments can be appeared at the decay of the intermediate state through the outer fission barrier. The quasi-fission fragments are different from the ones produced in the decay of the DNS on two fragments and related to the DIC or quasi-elastic processes. This is because the yield of the DIC or quasi-elastic fragments is concentrated around the incident nuclei, while the yield of the quasi-fission fragments is similar to the one for the compound nucleus fission.

The probabilities of compound nucleus formation for the systems $^{70}$Zn, and $^{78}$Ge + $^{208}$Pb are not described by Eqs. (12)-(13), because we should take into account the decay branches of the intermediate state. We consider the probability of compound nucleus formation in such case in next subsection.

\subsubsection{Expression for the probability of compound nucleus formation in the case of intermediate state}

The formation of compound nucleus in the case of the intermediate state is occurred by two steps. The first step is related to the formation of the intermediate state from the DNS, while the second one is linked to the decay of the intermediate state into the compound nucleus with the equilibrium shape. The intermediate state may decay into the compound nucleus, the quasi-fission fragments, and back to the DNS.

The consideration of the two-step process in the model is similar to the discussion of the sequential stages of the SHN formation used in Eq. (1). The intermediate state has taken place on the fusion way of the compound nucleus formation and has not any influence on the DNS path of the compound nucleus formation. As a result, the probability of compound nucleus formation in this case is determined as
\begin{eqnarray}
P(E,\ell)=\frac{\Gamma_{\rm CN}^{\rm DNS, f}(E,\ell) P_{\rm is}(E,\ell) +\Gamma_{\rm CN}^{\rm DNS}(E,\ell)}{\Gamma^{\rm tot}_{\rm CN}(E,\ell)} ,
\end{eqnarray}
where
\begin{eqnarray}
P_{\rm is}(E,\ell) = \frac{\Gamma^{\rm is}_{\rm CN}(E,\ell)}{\Gamma^{\rm is}_{\rm tot}(E,\ell)}
\end{eqnarray}
is the decay probability of the intermediate state into the compound nucleus. The widths presented in Eq. (17) have been discussed already, see Eqs. (12)-(13). Now we consider the widths appeared in Eq. (18).
\begin{eqnarray}
\Gamma^{\rm is}_{\rm tot}(E,\ell)=\Gamma^{\rm is}_{\rm CN}(E,\ell)+\Gamma_{\rm qf}^{\rm is}(E,\ell)+\Gamma_{\rm DNS}^{\rm is}(E,\ell)
\end{eqnarray}
is the total decay width of the intermediate state, $\Gamma^{\rm is}_{\rm CN}(E,\ell)$, $\Gamma_{\rm fiss}(E,\ell)$, and $\Gamma_{\rm DNS}^{\rm is}(E,\ell)$ are the decay widths of the intermediate state to the compound nucleus, the quasi-fission fragments, and the DNS, respectively. The width $\Gamma_{\rm qf}(E,\ell)$ is described the true quasi-fission process, which is related to the fission of the one-body nuclear system bypassing the formation of the compound nucleus with equilibrium shape.

The probability of compound nucleus formation decreases due to the decay of the intermediate state to the quasi-fission fragments or back to the DNS, because $P_{\rm is}(E,\ell) \leq 1$. Eq. (17) coincides with Eq. (12), when $P_{\rm is}(E,\ell)=1$.

The cross sections of the compound nucleus formation and the true quasi-fission are related to the corresponding decay branches of the intermediate state. Therefore, these cross sections can be defined, respectively, as
\begin{eqnarray}
\sigma_{\rm CN}(E)=\frac{\pi \hbar^2}{2\mu E} \sum_\ell (2l+1) T(E,\ell) P(E,\ell), \\
\sigma_{\rm qf}(E)=\frac{\pi \hbar^2}{2\mu E} \sum_\ell (2l+1) T(E,\ell) P_{\rm qf}(E,\ell).
\end{eqnarray}
Here
\begin{eqnarray}
P_{\rm qf}(E,\ell) =
\frac{\Gamma_{\rm CN}^{\rm DNS, f}(E,\ell)}{\Gamma^{\rm tot}_{\rm CN}(E,\ell)}
\times \frac{\Gamma_{\rm qf}^{\rm is}(E,\ell)}{\Gamma^{\rm is}_{\rm tot}(E,\ell))}\;\;
\end{eqnarray}
is the probability of the quasi-fission decay. The first factor in Eq. (22) is the probability of intermediate state formation, while the second one is the decay probability of the intermediate state to the quasi-fission fragments.

The number of the successive intermediate states $k$ may be more than one in the case of the very complex potential energy landscape. In such case the probability of compound nucleus formation is also determined by Eq. (17) in which the probability $P_{\rm is}(E,\ell)$ is substituted by the product of the decay probabilities of $k$ successive intermediate states $P_{\rm is\;1}(E,\ell) \cdot P_{\rm is\;2}(E,\ell) \cdot ... \cdot P_{{\rm is}\;k}(E,\ell)$. Eqs. (21)-(22) for the quasi-fission cross section should be also modified by respective way, because the quasi-fission fragments can be emitted at the decay of any intermediate state. The total number of corresponding parameters, which is needed for the reaction description, rises with the number of the intermediate states. Therefore, we consider in our model only the one intermediate state for reactions $^{70}$Zn and $^{78}$Ge + $^{208}$Pb.

The main decay channel of the superheavy compound nucleus is fission, consequently, the values of compound nucleus production cross sections are very close to the compound nucleus fission cross sections. The probabilities of formation of the compound nucleus fission fragments (12), (13) or quasi-fission fragments (22) are very small for the heavy cold-fusion systems. Therefore, the probabilities of these processes are strongly lower than the probability of the DIC fragments formed at the DNS decay, because it is necessary to form the compound nucleus or the intermediate state additionally to the DNS formation. This is strongly correlated to the experimental yields of near-symmetric fission or quasi-fission fragments and very asymmetric DIC or quasi-elastic fragments for various reactions \cite{120e,itkis,itkis-a,itkis-c}.

\subsubsection{The decay widths}

Eqs. (12)-(13), (17)-(19), and (22) include of the two types of the widths. The widths $\Gamma_{\rm CN}^{\rm DNS, f}(E,\ell)$, $\Gamma^{\rm is}_{\rm CN}(E,\ell)$, $\Gamma_{\rm qf}^{\rm is}(E,\ell)$, $\Gamma_{\rm DNS}^{\rm is}(E,\ell)$ are related to the one-body shape of nucleus, while the widths $\Gamma_{\rm CN}^{\rm DNS tr}(E,\ell)$, $\Gamma_{\rm DIC}^{\rm DNS}(E,\ell)$, $\Gamma_{\rm sph}^{\rm DNS}(E,\ell)$, $\Gamma_{\rm def}^{\rm DNS}(E,\ell)$ are linked to the two-body nuclear systems.

The widths linked to the various one-body shapes are determined as
\begin{eqnarray}
\Gamma_{\rm one-body}(E,\ell) = \frac{1}{\rho_{\rm in}(E)} \int_0^{E-B} d\varepsilon \; \rho_{A,\ell}(\varepsilon).
\end{eqnarray}
Here $\rho_{\rm in}(E)$ is the energy level density of the nuclear system in the initial state, $\rho_{A,\ell}(\varepsilon)$ is the energy level density of the nuclear system in the final state, $B$ is the height of the saddle point taken place on the way from the initial state to the final one, $\varepsilon$ is the excitation energy. The corresponding values of $E$, $\ell$ and $B$ should be applied at the calculation of the widths $\Gamma_{\rm CN}^{\rm DNS, f}(E,\ell)$, $\Gamma^{\rm is}_{\rm CN}(E,\ell)$, $\Gamma_{\rm qf}^{\rm is}(E,\ell)$, $\Gamma_{\rm DNS}^{\rm is}(E,\ell)$.

We use the back-shifted Fermi gas energy level density of nucleus with the excitation energy $\varepsilon$, $A$ nucleons and the angular momentum $J$ \cite{ripl3}, which is written as
\begin{eqnarray}
\rho_{A,J}(U) &=& \frac{(2 J+1)}{4 \sqrt{2 \pi} \sigma_J^3} \exp{ \left\{- [(J+1/2)/\sigma_J]^2/2 \right\}} \nonumber \\ &\times&
\frac{\sqrt{\pi}}{12 (a_{\rm dens} U^5)^{1/4}} \exp{\left[2 \sqrt{a_{\rm dens} U}\right]}.
\end{eqnarray}
Here
\begin{eqnarray}
U=\varepsilon-\delta= a_{\rm dens} T^2
\end{eqnarray}
is the back-shifted excitation energy, which is connected with the temperature $T$,
\begin{eqnarray}
\delta=12 n A^{-1/2}+0.173015
\end{eqnarray}
is the energy shift with $n= -1, 0$ and $1$ for odd-odd, odd-A, and even-even nuclei, respectively, $\sigma_J^2=(0.83 A^{0.26})^2$ is the spin cut-off parameter. The level density parameter depends on the excitation energy of the nucleus \cite{ist} and equals
\begin{eqnarray}
a_{\rm dens}=a_{\rm inf} [1 + (\delta_{\rm shell}/U) (1 -\exp{(-\gamma U)})],
\end{eqnarray}
where
\begin{eqnarray}
a_{\rm inf}=0.0722396 A + 0.195267 A^{2/3}
\end{eqnarray}
is the asymptotic level density parameter, $\gamma=0.410289/A^{1/3}$ is the damping parameter, $\delta_{\rm shell}$ is the phenomenological shell correction \cite{ripl3}. The value of phenomenological shell correction is determined as the difference $\delta_{\rm shell}=M_{\rm exp}-M_{\rm ld}$ \cite{ripl3}, where $M_{\rm exp}$ is the experimental value of the nuclear mass taken from Ref. \cite{be} and $M_{\rm ld}$ is the liquid drop component of the mass formula \cite{ms}. All parameter values used for the evaluation of the energy level density are taken from Ref. \cite{ripl3} without any changes. (Note the phenomenological shell corrections $\delta_{\rm shell}$ and $\delta E_i$, see Eq. (9), are the same physical sense. However, they are obtained using the different mass formulas for the calculation the liquid-drop contribution \cite{d2015,ripl3,ms}. The values of parameters $\delta_{\rm shell}$ and $\delta E_i$ are, respectively, linked to the values of other parameters of the energy level density and the nuclear part of the interaction potential. Therefore, we use for calculations of $\delta_{\rm shell}$ and $\delta E_i$ different expressions.) The values of shell corrections are very important for the properties of SHN \cite{d-shell}, therefore the influence of shell correction on the level density should be taken into account.

As we have pointed, the widths $\Gamma_{\rm CN}^{\rm DNS, tr}(E,\ell)$, $\Gamma_{\rm DIC}^{\rm DNS}(E,\ell)$, $\Gamma_{\rm sph}^{\rm DNS}(E,\ell)$, $\Gamma_{\rm def}^{\rm DNS}(E,\ell)$ are related to the DNS, which are consisted of two nuclei with various shapes and nucleon compositions. The width of the DNS built by the nuclei with numbers of nucleons $A_1$ and $A_2=A-A_1$, correspondingly, is written as
\begin{eqnarray}
\Gamma_{\rm DNS}(E,\ell) = \frac{1}{\rho_{\rm in}(E)} \int_0^{E-B_\ell} d\varepsilon \; \rho_{A_1,A_2}(\varepsilon,\ell) ,
\end{eqnarray}
where
\begin{eqnarray}
\rho_{A_1,A_2}(\varepsilon,\ell)= \int_0^{\varepsilon} d\varepsilon^\prime \; \rho_{A_1,0}(\varepsilon^\prime) \; \rho_{A_2,0}(\varepsilon-\varepsilon^\prime) .
\end{eqnarray}
Here $\rho_{\rm in}(E)$ is the energy level density of the nuclear system in the initial state. $\rho_{A_1,A_2}(\varepsilon,\ell)$ is the energy level density of the DNS, $\rho_{A_1,\ell}(\varepsilon)$ and $\rho_{A_2,\ell}(\varepsilon)$ are the energy level density of the nuclei with $A_1$ and $A_2$ nucleon in the final state, $B_\ell$ is the height of the saddle point taken place on the way from the initial state to the final one. We neglect by the transfer of the orbital moment of the DNS system into the orbital momenta of nuclei for the sake of simplicity.

The probabilities of the compound nucleus formation $P(E,\ell)$ and the decay of the intermediate state $P_{\rm is}(E,\ell)$ depend on the ratio of the decay widths into a specific states to the total decay width of the initial state. Therefore, these probabilities are independent on $\rho_{\rm in}(E)$.

We should define the barrier heights for the calculation of various decay widths and the probabilities. Let us consider the barriers for corresponding widths in details.

\subsubsection{The barrier heights for different processes}

The width $\Gamma_{\rm sph}^{\rm DNS}(E,\ell)$ depends on the barrier $B^{\rm fus}_\ell$. The value of $B^{\rm fus}_\ell$ was obtained at the calculation of the transmission probability $T(E,\ell)$, see Eq. (5). This barrier of the total potential energy of the spherical incident nuclei can be found using Eqs. (2), (6)-(10). Substituting value of $B^{\rm fus}_\ell$ to Eq. (29) we obtain $\Gamma_{\rm sph}^{\rm DNS}(E,\ell)$.

The width $\Gamma_{\rm def}^{\rm DNS}(E,\ell)$ connects to the barrier $B^{\rm fus}_{\ell,{\rm def}}$. This barrier is determined as the minimal values of the barrier of the total potential energy of deformed nuclei. The nucleon compositions of these deformed nuclei are the same as in the incident channel. The total potential energy of deformed nuclei is calculated in the framework of the approach developed in Refs. \cite{dms,ds17,dps}. Now we improve it by taking into account the realistic surface stiffness of interacting nuclei.

As shown in Refs. \cite{dp,dpil,dpil2,dpfus-a,dm,dm-a,dm-b} the axially-symmetric nuclei, which is elongated along the line connected the mass centers, have the lowest value of the barrier height. Therefore, we consider that the DNS decays preferably by such mutual orientation of the axially-symmetric nuclei. The other nucleus-nucleus configurations have higher values of the barrier. Consequently, such configurations have the lower values of the thermal excitation energy of the DNS and the smaller values of the statistical yield. As a result, such configurations we may be neglected.

The total potential energy of interacting deformed nuclei $V_{\rm DNS}(r,\ell,\{ \beta_{L1} \} ,\{ \beta_{L2} \})$ consists of the nuclear $V_{\rm N}(R,\{ \beta_{L1} \} ,\{ \beta_{L2} \})$, Coulomb $V_{\rm C}(R,\{ \beta_{L1} \} ,\{ \beta_{L2} \})$, and centrifugal $V_{\ell}(R,\{ \beta_{L1} \} ,\{ \beta_{L2} \})$ energies as well as the deformation energies $E_{{\rm def}i}(\{ \beta_{Li} \})$ of each nucleus. So, the total potential energy equals to
\begin{eqnarray}
V_{\rm DNS}(r,\ell,\{ \beta_{L1} \} ,\{ \beta_{L2} \}) = V_{\rm N}(r,\{ \beta_{L1} \} ,\{ \beta_{L2} \}) \nonumber \\ + V_{\rm C}(r,\{ \beta_{L1} \} ,\{ \beta_{L2} \}) + V_{\ell}(r,\{ \beta_{L1} \} ,\{ \beta_{L2} \}) \nonumber \\ + E_{{\rm def}1}(\{ \beta_{L1} \} )+ E_{{\rm def}2}(\{ \beta_{L2} \}),
\end{eqnarray}
where $\{ \beta_{Li} \}= \beta_{0i}, \beta_{1i}, \beta_{2i}$, $\beta_{3i}, \beta_{4i}$ is the set of the surface multipole deformation parameters of nucleus $i$, $i=1,2$. These deformation parameters are related to the surface radius of deformed nucleus
\begin{eqnarray}
R_i(\theta)=R_{0i} \left[ 1 + \sum_L \beta_{Li} Y_{L0}(\theta) \right],
\end{eqnarray}
where $R_{0i}$ is the radius of spherical nucleus $i$ and $Y_{L0}(\theta)$ is the spherical harmonic functions \cite{vmk}. The parameters $\beta_{0i}$ and $\beta_{1i}$ provide the volume conservation and non-movement of the position of the mass center for nucleus $i$. The values of the deformation parameters $\{ \beta_{L1} \} ,\{ \beta_{L2} \}$ are determined by the condition of the minima of the total interaction potential energy of these nuclei $V_{\rm DNS}(r,\ell,\{ \beta_{L1} \} ,\{ \beta_{L2} \})$ at given $r$. Note the contributions of higher multipole deformations $\beta_{L\geq 5}$ into the value of $V_{\rm DNS}(r,\ell,\{ \beta_{L1} \} ,\{ \beta_{L2} \})$ are negligible.

According to the proximity theorem \cite{derjaguin,prox}, the nuclear part of the interaction potential between deformed nuclei can be approximated as \cite{dms,ds17}
\begin{eqnarray}
V_{\rm N}(r,\{ \beta_{L1} \} ,\{ \beta_{L2} \})
 \approx S(\{ \beta_{L1} \}, \{ \beta_{L2} \}) \nonumber \\ \times V_{\rm N}^{\rm sph}(d(r,\{ \beta_{L1} \} ,\{ \beta_{L2} \})+R_{01}+R_{02})).
\end{eqnarray}
Here
\begin{eqnarray}
S(\{ \beta_{L1} \}, \{ \beta_{L2} \}) =\frac{ \frac{R_1(\pi/2)^2 R_2(\pi/2)^2}{R_1(\pi/2)^2 R_2(0)+R_2(\pi/2)^2 R_1(0)}}
{\frac{R_{01} R_{02}}{R_{01} + R_{02}}}
\end{eqnarray}
is the factor related to the modification of the strength of nuclear interaction of the deformed nuclei induced by the surface deformations, which is derived in Ref. \cite{dms},
\begin{eqnarray}
d(r,\{ \beta_{L1} \} ,\{ \beta_{L2} \})=r-R_1(0)-R_2(0)
\end{eqnarray}
is the smallest distance between the surfaces of deformed nuclei, which coincides to the distance between the surfaces of spherical nuclei. The potential $V_{\rm N}^{\rm sph}$ determines the nuclear part of the interaction between spherical nuclei, see Eqs. (6)-(10).

The expression of the Coulomb interaction of the two deformed arbitrary-oriented axial-symmetric nuclei is obtained by an expansion on the deformation parameters in Ref. \cite{dpil}. The accuracy of this expression is very high. The values of the Coulomb interaction of the two deformed arbitrary-oriented axial-symmetric nuclei evaluated by using the expression from Ref. \cite{dpil} and by the numerical calculations are very good agreed to each other \cite{iam}. Taking into account the considered orientation of axial-symmetric nuclei at the searching the value of the lowest barrier height we rewrite the expression from Ref. \cite{dpil} in a simple form
\begin{eqnarray}
V_{\rm C}(r) &=& \frac{Z_1 Z_2 e^2}{r} \{ 1 \nonumber \\ &+& \sum_{L \geq 1} \left[ f_{L1}(r,R_{01}) \beta_{L1} + f_{L1}(r,R_{02}) \beta_{L2} \right] \nonumber \\ &+& f_2(r,R_{01}) \beta_{21}^2 + f_2(r,R_{02}) \beta_{2 2}^2
\nonumber \\ &+& \left. f_3(r,R_{01},R_{02}) \beta_{2 1} \beta_{2 2} \right\},
\end{eqnarray}
where \begin{eqnarray}
f_{L1}(r,R_{0i}) = \frac{3R_{0i}^L}{2 \sqrt{\pi(2L+1)}r^L}, \\
f_2(r,R_{0i}) = \frac{3 R_{0i}^2}{7\pi r^2}
+ \frac{9 R_{0i}^4}{14 \pi r^4},
\\
f_3(r,R_{01},R_{02}) =
\frac{27 R_{01}^2 R_{02}^2}{10 \pi r^4} .
\end{eqnarray}
This expression is taken into account linear and quadratic terms on the quadrupole deformation parameters and linear terms of high-multipolarity deformation parameters. The volume correction, which is appeared in the second order of the quadrupole deformation parameter and important for heavy systems, is taken into account in this expression.

The nuclei formed DNS after penetration the fusion barrier are excited. Therefore, the momentum of inertia of the DNS can be well approximated in the framework of the solid-state model. The centrifugal potential energy of DNS nuclei is
\begin{eqnarray}
V_{\ell}(r,\{ \beta_{L1} \} ,\{ \beta_{L2} \}) = \frac{\hbar^2 \ell(\ell+1)}{2 (\mu r^2+J_1+J_2)},
\end{eqnarray}
where \begin{eqnarray}
J_i=(2/5)m_n R_{0i}^2 A_i (1+\sqrt{5/(16 \pi)} \beta_{2i})
\end{eqnarray}
is the momentum of inertia of nucleus $i$, and $m_n$ is the mass of nucleon. Here we take into account only quadrupole deformation, because the contribution of higher multipolarities into the momentum of inertia is negligible.

The incident nuclei participated in the cold-fusion reactions have the spherical equilibrium shape. The nuclei involved to the DNS evolution are deforming due to interaction between them. The deformation energy of the nucleus induced by a deviation from the spherical shape consists of the surface and Coulomb contributions. This energy in the liquid-drop approximation \cite{bormot} is given as
\begin{eqnarray}
E_{{\rm def} i }^{\rm ld}(\{ \beta_{Li} \}) = \sum_{L=2}^4 C^{\rm ld}_{LA_iZ_i} \frac{\beta_{Li}^2}{2},
\end{eqnarray}
where
\begin{eqnarray}
C^{\rm ld}_{LA_iZ_i}= \frac{(L-1)(L+2) b_{\rm surf} A^{2/3}_i}{4 \pi} - \frac{3(L-1) e^2 Z^2_i}{2\pi(2L+1)R_{0i}}
\end{eqnarray}
is the surface stiffness coefficient obtained in the liquid-drop approximation, and $b_{\rm surf}$ is the surface coefficient of the mass formula \cite{msis}.

We can also evaluate the realistic deformation energy of a nucleus at small surface deformations in the framework of the shell correction method \cite{s,s1,s2,s3} and approximate the dependence of this energy on the deformation parameters by
\begin{eqnarray}
E_{{\rm def} i }^{\rm sc}(\{ \beta_{Li} \}) = \sum_{L=2}^4 C^{\rm sc}_{LA_iZ_i} \frac{\beta_{Li}^2}{2}.
\end{eqnarray}
Here $C^{\rm sc}_{LA_iZ_i}$ is the total surface stiffness coefficient obtained with the shell correction method. Using the shell-correction method we can split both the deformation energy and the stiffness coefficient on the shell-correction and liquid-drop parts
\begin{eqnarray}
E_{{\rm def} i }^{\rm sc}(\{ \beta_{Li} \}) = E_{{\rm def} i }^{\rm shell}(\{ \beta_{Li} \})+E_{{\rm def} i }^{\rm ld}(\{ \beta_{Li} \}) \nonumber \\
= \sum_{L=2}^4 [C^{\rm shell}_{LA_iZ_i}+C^{\rm ld}_{LA_iZ_i}] \frac{\beta_{Li}^2}{2} \nonumber \\ = \sum_{L=2}^4 \left[ \left(\frac{C^{\rm sc}_{LA_iZ_i}}{C^{\rm ld}_{LA_iZ_i}}-1\right)+1 \right] C^{\rm ld}_{LA_iZ_i} \frac{\beta_{Li}^2}{2} .
\end{eqnarray}

The deformation energy of a nucleus at small surface deformations can be obtained in the harmonic oscillator model too \cite{bormot,wong68}. In this model the deformation energy of a nucleus is described as
\begin{eqnarray}
E_{{\rm def} i }^{\rm ho}(\{ \beta_{Li} \}) = \sum_{L=2}^4 C^{\rm ho}_{LA_iZ_i} \frac{\beta_{Li}^2}{2}.
\end{eqnarray}
Here $C^{\rm ho}_{LA_iZ_i}$ is the surface stiffness coefficient in the harmonic oscillator model, which is connected to the energy ${\cal E}_{LA_iZ_i}$ and the total zero-point amplitude $\beta_{LA_iZ_i}^0$ of the surface oscillations (or the transition probability for exciting the surface oscillations $B(E,0\rightarrow L)$) \cite{bormot,wong68}
\begin{eqnarray}
C^{\rm ho}_{LA_iZ_i} &=& \frac{(2L+1){\cal E}_{LA_iZ_i}}{2 (\beta_{LA_iZ_i}^0)^2} \nonumber \\ &=& \left(\frac{3ZeR^L}{4 \pi} \right)^2\frac{(2L+1){\cal E}_{LA_iZ_i}}{2 B(E,0\rightarrow L)}.
\end{eqnarray}
The known experimental values of ${\cal E}_{LA_iZ_i}$, $\beta_{LA_iZ_i}^0$, and/or $B(E,0\rightarrow L)$ for nuclei are tabulated for $L=2$ and 3 in Refs. \cite{be2,be3}. Note the coupling of the incident channel with the low-energy surface vibration channels is often taken into account in the framework of the harmonic oscillator approach at description of various heavy-ion reactions \cite{fl,dhrs,motstef,den-subfus}. The characteristics of heavy-ion reactions depend strongly on the properties of the surface vibrations.

The harmonic oscillator $C^{\rm ho}_{LA_iZ_i}$ and shell-correction $C^{\rm sc}_{LA_iZ_i}$ values of the surface stiffness parameters should be close to each other. Therefore, we rewrite Eq. (45) in the form
\begin{eqnarray}
E_{{\rm def} i }^{\rm sc}(\{ \beta_{Li} \}) = \sum_{L=2}^4 \left[ \left(\frac{C^{\rm ho}_{LA_iZ_i}}{C^{\rm ld}_{LA_iZ_i}} -1 \right) +1 \right] C^{\rm ld}_{LA_iZ_i}\frac{\beta_{Li}^2}{2}. \;\;\;
\end{eqnarray}
This expression for deformation energy is useful for further application, because using experimental values of ${\cal E}_{LA_iZ_i}$, $\beta^0_{LA_iZ_i}$ we find the values of the ratio $C^{\rm ho}_{LA_iZ_i}/C^{\rm ld}_{LA_iZ_i}$, see in Table 2. We put $C^{\rm ho}_{LA_iZ_i} / C^{\rm ld}_{LA_iZ_i}=1$, if the experimental data for evaluation of $C^{\rm ho}_{LA_iZ_i}$ are unknown.

The values of the ratio $C^{\rm ho}_{LA_iZ_i} / C^{\rm ld}_{LA_iZ_i}$ for $L=2,3$ presented in Table 2 have an irregular behaviour from one nucleus to another, see also in Refs. \cite{bormot,wong68}. This ratio very strongly deviates from 1 near the magic nuclei. Nucleus $^{208}$Pb is very stiff for the surface quadrupole and octupole distortions, because $C^{\rm ho}_{LA_iZ_i} / C^{\rm ld}_{LA_iZ_i} \gg 1$ for $L=2,3$. In contrast to this, nuclei $^{58}$Fe and $^{78}$Ge are soft for the surface quadrupole distortions, because $C^{\rm ho}_{LA_iZ_i} / C^{\rm ld}_{LA_iZ_i} \ll 1$ for $L=2$. These nuclei are well deformed during the DNS decay.

\begin{table}
\caption {\label{tab2}The ratio $C^{\rm ho}_{LA_iZ_i} / C^{\rm ld}_{LA_iZ_i}$ obtained using the experimental properties of the low-energy surface vibrational states with multiplicities $L=2$ \cite{be2} and $L=3$ \cite{be3}. We put $C^{\rm ho}_{LA_iZ_i} / C^{\rm ld}_{LA_iZ_i}=1$ in the case of unknown the experimental properties of the low-energy surface vibrational states for some multiplicity and nucleus. }
\begin{center}
\begin{tabular}{|c|ccc|}
\hline
Nucleus & \multicolumn{3}{c|}{ $C^{\rm ho}_{LA_iZ_i} / C^{\rm ld}_{LA_iZ_i}$} \\
 & $L=2$ & $L=3$ & $L=4$ \\
\hline
$^{50}$Ti & 2.03 & 1.46 & 1 \\
$^{52}$Cr & 12 & 3.15 & 1 \\
$^{54}$Cr & 0.45 & 1 & 1 \\
$^{58}$Fe & 0.36 & 2.0 & 1 \\
$^{64}$Ni & 1.46 & 1.36 & 1 \\
$^{70}$Zn & 0.54 & 0.56 & 1 \\
$^{78}$Ge & 0.33 & 1 & 1 \\
$^{208}$Pb & 44.9 & 2.2 & 1 \\
\hline
\end{tabular}
\end{center}
\end{table}

The typical values of excitation energy of DNS, formed by incident nuclei in cold-fusion reactions with 1--3 evaporated neutrons are 15--40 MeV. The amplitudes of shell correction energy at such excitation energy are approximately reduced 2--4 times \cite{ds,ach,ach-b,bq,dah,lpc,snp,pnsk,pnsk-a,zp}. We expect a similar effect for the value of the stiffness parameter, which should approach to the hydrodynamical one at high excitation energies.

Moreover, the single-particle spectra of nuclei near the contact point became more homogeneous due to a level splitting and shifting induced by the nucleus-nucleus interaction. This leads to a reduction of the amplitudes of the shell correction energies in interacting nuclei, see also Eq. (8). Consequently, the values of realistic surface stiffness coefficient of nuclei should approach to the liquid-drop one at small distances between them due to the nucleus-nucleus interaction.

Taking into account the excitation energy and nucleus-nucleus interaction effects on the shell correction energies we modify Eq. (48) as
\begin{eqnarray}
E_{{\rm def} i }(\{ \beta_{Li} \}) = \sum_{L=2}^4 \left[ \left(\frac{C^{\rm ho}_{LA_iZ_i}}{C^{\rm ld}_{LA_iZ_i}} -1\right)k_{LA_iZ_i} +1\right] \nonumber \\ \times \frac{C^{\rm ld}_{LA_iZ_i} \beta_{Li}^2}{2}.
\end{eqnarray}
Here $k_{LA_iZ_i} \approx 0.1$ is the parameter, which is described the attenuation of the shell-correction effect on the surface stiffness coefficient of the incident nuclei formed the DNS in the cold-fusion reactions. If $C^{\rm ho}_{LA_iZ_i}=C^{\rm ld}_{LA_iZ_i}$ then the deformation energy is determined by the liquid-drop properties and independent on $k_{LA_iZ_i}$. Note, the deformation energy is only defined by the liquid drop properties in the framework of various versions of the DNS model of the SHN production \cite{dns1,dns1c,dns1aa,dns1aa1,dns1aa2,dns1aa4,dns2a,adamian,dns3,dns31,dns32,dns33,dns3a,dns4b,dns4c,dns4d,dns5}.

The double-magic target nucleus $^{208}$Pb and magic or close to magic projectile nuclei are involved in the incident channel of the cold-fusion reactions. Therefore, we should take into account the realistic surface stiffness of nuclei at calculation of the width $\Gamma_{\rm def}^{\rm DNS}(E,\ell)$. The width $\Gamma_{\rm def}^{\rm DNS}(E,\ell)$ is linked to $B^{\rm DNS}_{\ell,{\rm def}}$, which is calculated with the help of Eqs. (31)-(41), (43), (47), (49). The value of $B^{\rm DNS}_{\ell,{\rm def}}$ rises with a rising $C^{\rm ho}_{2A_i82}/C^{\rm ld}_{2A_i82}$, because the barrier has taken place at smaller values of the deformation parameter of nuclei. The using very stiff nuclei in the cold fusion reaction leads to a higher value of $B^{\rm DNS}_{\ell,{\rm def}}$ and, as a result, a smaller value of $\Gamma_{\rm def}^{\rm DNS}(E,\ell)$. This leads to the increasing of the probability of compound nucleus formation $P(E,\ell)$ described by Eq. (12). In the contrary, fusion reactions between soft nuclei have a smaller value of $B^{\rm DNS}_{\ell,{\rm def}}$ and, as a result, higher value of $\Gamma_{\rm def}^{\rm DNS}(E,\ell)$ and smaller value of $P(E,\ell)$.

The correlation between the surface stiffness of incident nuclei and the production cross sections is clearly observed experimentally. For example, the values of $C^{\rm ho}_{2A_i82}/C^{\rm ld}_{2A_i82}$ for nuclei $^{208,206,204}$Pb obtained using data from Ref. \cite{be2} are, respectively, close to 45, 25, 17. The values of cross-section maxima for reactions $^{208,206,204}$Pb($^{48}$Ca,$2n)^{254,252,250}$No are $3\cdot 10^6, 4 \cdot 10^5, 7\cdot 10^3$ b \cite{102jinr,102jinr1}, correspondingly. So, we clearly see that reactions with more stiff target nuclei have higher values of the SHN production cross section. (Note, the other effects may also contribute into the cross-section values.) The surface stiffness effect may be also significant for the synthesis of SHN with $Z> 118$ in hot fusion reactions, when the stiff projectile $^{48}$Ca is substituted by more soft the one as $^{50}$Ti or similar.

Let us consider the other barriers related corresponding decay widths used in our model. The barrier $B_{\ell,{\rm DIC}}^{\rm DNS}=B_{0,\ell,{\rm DIC}}^{\rm DNS}+Q_{\rm tr}$ is defined as the barrier between deformed contacting nuclei formed after the nucleon transfer from the heavy nucleus to the light one, where $Q_{\rm tr}$ is the transfer reaction Q-value evaluated with the help of an atomic mass table \cite{be}. This barrier is taken place at evolution of initial DNS system to the more symmetric one. The DNS after passing the barrier $B_{\ell,{\rm DIC}}^{\rm DNS}$ can decay in the two scattered nuclei with new nucleon composition or the nucleon exchange between nuclei can continue further. The interaction potential energy of touching nuclei after nucleon exchange $B_{0,\ell,{\rm DIC}}^{\rm DNS}$ is calculated the similar way as $B^{\rm DNS}_{\ell,{\rm def}}$. The barrier $B_{\ell,{\rm DIC}}^{\rm DNS}$ is the minimal value of among the barriers related to various nucleon transfer paths from the incident DNS to the more symmetric the one. Substituting the obtained value of the barrier into Eq. (29) we can find the width $\Gamma_{\rm DIC}^{\rm DNS}(E,\ell)$.

The compound nucleus formation using the DNS path is related to the barrier $B_{\ell,{\rm CN}}^{\rm DNS, tr}$, which takes place at nucleons transfer from the light nucleus to the heavy one. The values of $B_{\ell,{\rm CN}}^{\rm DNS, tr}$ for every system formed at along various multi-nucleon transfer paths is evaluated similarly to $B_{\ell,{\rm DIC}}^{\rm DNS}$. The surface deformations of both nuclei are also taken into account. The barrier $B_{\ell,{\rm CN}}^{\rm DNS, tr}$ is the minimal value among the barriers related to various paths from the DNS formed by incident nuclei to the compound nucleus. The width $\Gamma_{\rm CN}^{\rm DNS, tr}(E,\ell)$ is calculated substituting the value of $B_{\ell,{\rm CN}}^{\rm DNS, tr}$ into Eq. (29).

The width $\Gamma_{\rm CN}^{\rm DNS, f}(E,\ell)$ is related the barrier
\begin{eqnarray}
B_{\ell,{\rm CN}}^{\rm DNS, f} = B_{0,{\rm CN}}^{\rm DNS, f} +\frac{\hbar^2 \ell(\ell+1)}{2 J_{\rm CN}^{\rm fus}} +Q_{\rm CN}.
\end{eqnarray}
Here $B_{0,{\rm CN}}^{\rm DNS, f}$ the height of a corresponding saddle point evaluated relatively the ground state of the compound nucleus using the potential energy surface presented in Figs. 1--2, $J_{\rm CN}^{\rm fus}$ is the momentum of inertia of compound nucleus at the saddle point, and $Q_{\rm CN}$ is the Q-value of the compound nucleus formation evaluated with the help of an atomic mass table \cite{be}. The width $\Gamma_{\rm CN}^{\rm DNS, f}(E,\ell)$ is obtained using Eq. (23) and the value $B_{\ell,{\rm CN}}^{\rm DNS, f}$.

The widths related to the intermediate state can be found similar way as the width $\Gamma_{\rm CN}^{\rm DNS, f}(E,\ell)$.

After obtaining the values of all widths we can determine the probability of compound nucleus formation. Now we can determine the survival probability of compound nucleus.

\subsection{The survival probability of compound nucleus}

The survival probability of the compound nucleus formed in the cold-fusion reaction relates to the competition between the evaporation of $x$ neutrons and the fission. It can be approximated by expression
\begin{eqnarray}
W^{xn}(E,\ell) = P_{xn}(E_{{\rm CN}\ell}^*) \frac{\Gamma_{1n}(E^*_1,\ell)}{\Gamma_{1n}(E^*_{1n},\ell)+\Gamma_{\rm f}(E^*_1,\ell)} \nonumber \\ \times \frac{\Gamma_{2n}^{A-1}(E^*_2,\ell)}{\Gamma_{2n}^{A-1}(E^*_2,\ell)+\Gamma_{\rm f}^{A-1}(E^*_2,\ell)} \times ... \nonumber \\ \times \frac{\Gamma_{xn}^{A-x+1}(E^*_x,\ell)}{\Gamma_{xn}^{A-x+1}(E^*_x,\ell)+\Gamma_{\rm f}^{A-x+1}(E^*_x,\ell)}.
\end{eqnarray}
Here $P_{xn}(E^*)$ is the realization probability of the $xn$-evaporation channel \cite{jack}, $E^*_{{\rm CN}\ell}= E-Q_{\rm CN}-\hbar^2 \ell(\ell+1)/(2J_{\rm gs})$, $J_{\rm gs}$ is the ground-state momentum of inertia. $E^*_1=E-Q_{\rm CN}$ is the excitation energy of the compound nucleus formed in the heavy-ion fusion reaction.
$\Gamma_{yn}^{A-y+1}(E^*_y,\ell)$ and $\Gamma_{\rm f}^{A-y+1}(E^*_y,\ell))$ are, respectively, the width of neutron emission and the fission width of the compound nucleus formed after emission of $(y-1)$ neutrons. $E^*_y=E^*_{y-1}-B_{n,y-1} - 2T_{y-1}$ is the excitation energy before evaporation of the $y$-th neutron, where $B_{n,y-1}$ is the separation energy of the $(y-1)$-th neutron. $T_{y-1}$ is the temperature of the compound nucleus after evaporation of $(y-1)$ neutrons and obtained from $E^*_{y-1}= a_{\rm dens} T_{y-1}^2$, where $a_{\rm dens}$ is defined by Eqs. (27)-(28).

The width of neutron emission from the nucleus with $A$ nucleons is given as \cite{vh}
\begin{eqnarray}
\Gamma_n(E^*,\ell) = \frac{g_n m_n R_n^2}{\pi\hbar^2 \rho_{A,\ell}(E^*)} \int_0^{E^*-B_n} d\varepsilon \; \varepsilon \nonumber \\ \times \rho_{A-1,\ell}(E^*-B_n-\varepsilon ,\ell), \;\;
\end{eqnarray}
where $B_n$ is the neutron separation energy from the nucleus, $\rho_{A,\ell}(E^*)$ and $\rho_{A-1,\ell}(E^*)$ are, correspondingly, the energy level densities of the compound nuclei before and after neutron emission, $g_n$ is the neutron intrinsic spin degeneracy, and $R_n$ is the radius of the neutron-nucleus interaction.

The fission width of the nucleus depends on the fission barrier height, which is consisted of the liquid-drop and shell-correction contribution in the Strutinsky shell correction prescription \cite{s,s1,s2,s3}. The excitation energy of a compound nucleus formed in cold-fusion reactions $E^*$ belongs to the range 10 to 25 MeV, therefore $T \lesssim 1$ MeV. The liquid-drop part of the fission barrier weakly depends on temperature at $T \lesssim 2$ MeV \cite{ds,pvb,bgh,gsb}. As a result, the temperature dependence of the shell correction contribution \cite{ds,ach,ach-b,bq,lpc,dh,dp,dah} induces the temperature dependence of the fission barrier of SHN.

The exponential reduction of the fission barrier of SHN with thermal excitation energy is obtained in the framework of the finite-temperature self-consistent Hartree-Fock+BCS model with Skyrme force in Refs. \cite{pnsk,pnsk-a,snp,zp}. The exponential dependence of the fission barrier height is also used in Refs. \cite{dh,dns1,dns1c,dns1aa,dns1aa1,dns1aa2,dns1aa4,dns2a,dns3,dns31,dns32,dns33,dns3a,dns4b,dns4c,dns4d,param3,ds,ds2019}. We also consider the exponentially decreasing of the fission barrier with the excitation energy and define the fission barrier of excited rotating nuclei as
\begin{eqnarray}
B_{\rm f}(\varepsilon,\ell) = B_{\rm f}^{\rm ld}+B_{\rm f}^{\rm sh} e^{-\gamma_D \varepsilon} + \frac{\hbar^2 \ell(\ell+1)}{2} \left[ \frac{1}{J_{\rm s}} - \frac{1}{J_{\rm gs}} \right]. \; \;
\end{eqnarray}
Here $B_{\rm f}^{\rm ld}$ and $B_{\rm f}^{\rm sh}(\varepsilon)$ are the liquid-drop and shell-correction contribution of the fission barrier, $\gamma_D $ is the damping parameter \cite{dp,ds,ds2019,dh,snp,pnsk,pnsk-a}. The last line of equation describes the rotational contribution to the barrier. $J_{\rm gs (s)}= m_n (2/5)R_0^2 A(1+\sqrt{5/(16 \pi)} \beta_{\rm gs(s)})$ and $\beta_{\rm gs(s)}$ are, respectively, the ground-state (fission saddle-point) momentum of inertia and quadrupole deformations of the compound nucleus.

The dependence of the fission barrier of SHN on the excitation energy should be taken into account in the evaluation of the survival probability. The Bohr-Wheeler expression for the fission width \cite{bw,vh} is obtained in the transition state approach with the fission barrier independent of the excitation energy. This expression is not consistent with the barrier dependent on the excitation energy \cite{ssww,ssww-a}.

The number of states over barrier increases with the thermal reduction the fission barrier height. Taking into account both the dependence of the fission barrier on the excitation energy and the rising of the number of states over the energy-dependent fission barrier we derived new expression for the fission width in the form \cite{ds}
\begin{eqnarray}
\Gamma_{\rm f}(E^*,\ell)=\frac{2}{2\pi \rho_{A,\ell}(E^*)} \int_0^{\varepsilon_{\rm max}} d\varepsilon \frac{\rho_{A,\ell}(\varepsilon)}{N_{\rm tot}} N_{\rm saddle}(\varepsilon).
\end{eqnarray}
Here the ratio $\rho(\varepsilon)/N_{\rm tot}$ is the probability to find the fissioning nucleus with the intrinsic thermal excitation energy $\varepsilon$ in the fission transition state, $N_{\rm tot}= \int_0^{\varepsilon_{\rm max}} d\varepsilon \rho_{A,\ell}(\varepsilon)$ is the total number of states available for fission in the case of the energy-dependent fission barrier, $N_{\rm saddle}(\varepsilon) = \int^{E-B_{\rm f}(\varepsilon)}_{\varepsilon} de \rho_{A,\ell}(e)$ is the number of states available for the fission at $\varepsilon$ and the barrier value $B_{\rm f}(\varepsilon)$. $\varepsilon_{\rm max}$ is the maximum value of the intrinsic thermal excitation energy of nucleus at the saddle point, which is determined as the solution of the equation
\begin{eqnarray}
\varepsilon_{\rm max} + B_{\rm f}(\varepsilon_{\rm max},\ell)= E^*.
\end{eqnarray}
This equation is related to the energy conservation law, i.e. the sum of thermal $\varepsilon_{\rm max} $ and potential $B_{\rm f}(\varepsilon_{\rm max},\ell)$ energies at the saddle point equals to the total excitation energy $E^*$.

The difference between the Bohr-Wheeler fission width \cite{bw} and $\Gamma_{\rm fiss}(E^*,\ell)$ is discussed in Refs. \cite{ds2019,ds}. Note, $\Gamma_{\rm f}(E^*,\ell)$ is equal to the Bohr-Wheeler fission width in the case of the energy-independent fission barrier \cite{ds}.

\section{Discussion}

In this section we compare the theoretical values of the SHN production cross sections obtained in the framework of our model for various cold fusion reactions with the experimental ones. At the beginning we like to point some important experimental features of the cross-section data of the SHN production.

The experimental cross sections for reaction $^{208}$Pb($^{64}$Ni,$n)^{271}$Ds are measured in Gesellschaft fur Schwerionenforschung (GSI) \cite{h,108gsi}, Lawrence Berkeley National Laboratory (LBNL) \cite{110lbnl,folden}, and Institute of Physical and Chemical Research (RIKEN) \cite{110riken}. Unfortunately, the collision energies and the values of the cross section in the maxima obtained in different laboratories for this reaction are different. The difference between the lowest \cite{h,108gsi} and highest \cite{110riken} collision energies of the maxima of the cross sections for reaction $^{208}$Pb($^{64}$Ni,$n)^{271}$Ds obtained in different experiments is close to 5 MeV (see also Fig. 7 and Ref. \cite{folden}). This value of energy difference is very large, because the cold fusion reaction is usually taken place at sub-barrier collision energies. The values of the sub-barrier fusion cross section may be changed on several orders at the variation of the collision energy of 5 MeV. Such changes of the sub-barrier fusion cross section on collision energy are typical for light and medium heavy-ion systems, see, for example, Refs. \cite{dp,fl,dpil2,dhrs,motstef,den-subfus}. Similar dependence of the capture process should be in the case of cold fusion reactions too.

The targets used in experiments for the SHN synthesis are thick. The reaction of SHN production may take place when the projectile is just entering into the target or just before the projectile escaping the target. The beam loss energy in the target is close to 3--4 MeV \cite{folden}. As a rule, the collision energies in the middle of the target or the laboratory beam energies are pointed in the experimental works. In these cases, the reaction may take place at energies on approximately $\pm 2$ MeV or $4$ MeV lower from the experimentally pointed ones.

The maximal values of the cold-fusion cross section measured in different laboratories are strongly varied too. For example, the maximal values of the cross section for reaction $^{208}$Pb($^{50}$Ti,$n)^{257}$Rf presented in Refs. \cite{104gsi,104gsi-a,104lbnl} differ about three times, (see also Fig. 3). Note that the number of reaction events is relatively high for this reaction. Due to this the statistical errors are low in both experiments. In contrast to this, 1 or 2 reaction events are only obtained in experiments for the production of very heavy SHN in cold-fusion reactions. Therefore, the experimental errors of the cross section are very high, (see, for example, Fig. 8).

Due to these reasons, the exact values of the heavy-ion collision energies at which the SHN form in experiments and the cross-section values have not been well-defined now. Consequently, this is not a sense to describe the cross-section of the SHN production precisely. Therefore, we use a "soft" criterion of the agreement between the experimental data and the theoretical calculations. This criterion assumes that we describe both the maximal value of the cross section for some experimental measurement with 50\% accuracy and the energy position of the maximum of the cross section with a precision of several MeVs. Moreover, we see the more physical sense in the smaller changes of the fitting parameters for the nearest reactions, than in better agreement with the data. Below we will point, when the fitting parameters drastically changed for the nearest reactions and discuss the reason of these changes.

\subsection{Reaction $^{208}$Pb($^{50}$Ti,$xn)^{258-x}$Rf}

\begin{figure}
\includegraphics[width=8.5cm]{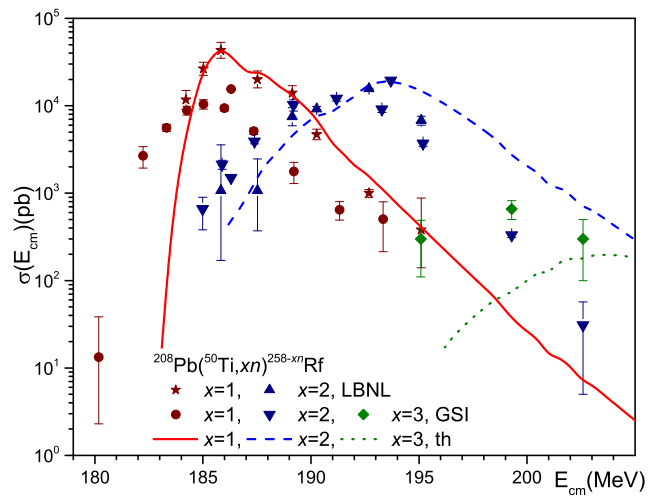}
\caption{\label{fig3} A comparison of our theoretical calculations of the cross sections for reactions $^{208}$Pb($^{50}$Ti,$xn)^{258-x}$Rf with $x=1,2$ and 3 with available experimental data. The cross sections for reactions $^{208}$Pb($^{50}$Ti,$xn)^{258-x}$Rf with $x=1$ and 2 are measured in Refs. \cite{104gsi,104gsi-a} (GSI) and \cite{104lbnl} (LBNL) and the cross sections for reaction $^{208}$Pb($^{50}$Ti,$3n)^{255}$Rf are obtained in Ref. \cite{104gsi,104gsi-a} (GSI).}
\end{figure}

The values of the cross sections for reactions $^{208}$Pb($^{50}$Ti,$xn)^{258-x}$Rf with $x=1$ and 2 are measured in GSI \cite{104gsi,104gsi-a} and LBNL \cite{104lbnl}. The experimental cross sections for reaction $^{208}$Pb($^{50}$Ti,$3n)^{255}$Rf are only obtained in GSI \cite{104gsi,104gsi-a}. We calculate the cross sections for these reactions in the framework of our model and compare the obtained values with the experimental data in Fig. 3. Our results agree well with the LBNL data \cite{104lbnl}, but the GSI data shift to the lower collision energies relatively both the LBNL data and our results, see Fig. 3. Taking into account the "soft" criterium we conclude that the experimental cross sections for reactions $^{208}$Pb($^{50}$Ti,$xn)^{258-x}$Rf with $x=1 \div 3$ are well described in the framework of our model.

The parameters of the model used in the calculation of reactions $^{208}$Pb($^{50}$Ti,$xn)^{258-x}$Rf are presented in Tables 2--4. Note, these values of parameters are not unique. It is also possible to describe the data by choosing slightly other values of parameters.

The starting values of the fission barrier $B_{\rm f}(0,0)$ and the compound nucleus formation barrier $B_{\rm 0,CN}^{\rm DNS f}$ are extracted from Fig. 1. However, the final values of the parameters presented in Tables 3--4 are chosen by the fine fitting of the experimental data. The values of these barriers given in Tables are close to the ones extracted in Fig. 1. The ground state $\beta_{\rm gs}$ and saddle point $\beta_{\rm sp}$ deformations of fissioning nuclei are taken from Fig. 1.
\\

\begin{table}
\caption{\label{tab3}The liquid-drop $B_{\rm f}^{\rm ld}$ and shell $B_{\rm f}^{\rm sh}$ contributions of the fission barriers $B_{\rm f}^0=B_{\rm f}^{\rm ld}+B_{\rm f}^{\rm sh}$ of nuclei, the ground state $\beta_{\rm gs}$ and saddle point $\beta_{\rm sp}$ deformations of fissioning nuclei, and the damping parameter of the fission barrier $\gamma_D$. The fission barrier values $B_{\rm f}$ obtained in Refs. \cite{etfsi,mol} are also presented. The values of barriers are given relatively the ground-state energy of the compound nucleus in MeV. The values of $\gamma_D$ are presented in MeV$^{-1}$.}
\begin{center}
\begin{tabular}{ccccccccc}
\hline
Nucleus & $B_{\rm f}^{\rm ld}$ & $B_{\rm f}^{\rm sh}$ & $B_{\rm f}^0$ & $B_{\rm f}^{[127]}$ & $B_{\rm f}^{[135]}$ & $\beta_{\rm gs}$ & $\beta_{\rm sp}$ & $\gamma_D$ \\
\hline
$^{258}$Rf & 0.5 & 6.3 & 6.8 & 5.0 & 5.65 & 0.2 & 0.4 & 0.105 \\
$^{257}$Rf & 0.5 & 6.1 & 6.6 & 5.6 & 6.02 & 0.2 & 0.4 & 0.105 \\
$^{256}$Rf & 0.5 & 6.5 & 7.0 & 5.3 & 6.26 & 0.2 & 0.4 & 0.105 \\
$^{262}$Sg & 0.4 & 3.7 & 4.1 & 4.3 & 5.91 & 0.2 & 0.4 & 0.07 \\
$^{261}$Sg & 0.4 & 3.3 & 3.7 & 4.7 & 5.88 & 0.2 & 0.4 & 0.07 \\
$^{260}$Sg & 0.4 & 4.8 & 5.2 & 4.6 & 5.84 & 0.2 & 0.4 & 0.11 \\
$^{259}$Sg & 0.3 & 3.0 & 3.3 & 4.9 & 5.82 & 0.2 & 0.4 & 0.11 \\
$^{266}$Hs & 0.5 & 5.7 & 6.3 & 3.5 & 6.26 & 0.2 & 0.4 & 0.10 \\
$^{265}$Hs & 0.5 & 4.2 & 4.7 & 3.5 & 6.26 & 0.2 & 0.4 & 0.10 \\
$^{272}$Ds & 0.4 & 3.7 & 4.1 & 2.2 & 7.31 & 0.2 & 0.4 & 0.04 \\
$^{271}$Ds & 0.3 & 3.5 & 3.8 & 2.2 & 6.92 & 0.2 & 0.4 & 0.04 \\
$^{278}$Cn & 0.2 & 2.3 & 2.5 & 1.9 & 5.99 & 0.0 & 0.3 & 0.04 \\
$^{277}$Cn & 0.3 & 2.6 & 2.9 & 2.0 & 6.36 & 0.0 & 0.3 & 0.04 \\
$^{286}$Fl & 0.4 & 4.2 & 4.6 & 4.1 & 9.00 & 0.0 & 0.3 & 0.05 \\
$^{285}$Fl & 0.4 & 4.0 & 4.4 & 2.7 & 8.82 & 0.0 & 0.3 & 0.05 \\
\hline
\end{tabular}
\end{center}
\end{table}

The experimental information on the fission barrier height in SHN is very poor. The values of the fission barrier obtained in various models \cite{sob,sob1,etfsi,d-uhe,mol,snp,pnsk,pnsk-a,kowal,kowal-a,kjs,afanas,afanas-a,agbemava,warga} for SHN are very different. The difference reaches more than 100\% in some cases \cite{agbemava}. The reasons of uncertainties in the predictions of the fission barrier heights in SHN have been discussed in Ref. \cite{agbemava}. Therefore, it is difficult to prefer any model for the barrier calculation. Nevertheless, we present the fission barrier values $B_{\rm f}$ obtained in Refs. \cite{etfsi,mol} in Table 3 for a comparison. The values of barriers $B_{\rm f}(0,0)=B_{\rm f}^{\rm ld}+B_{\rm f}^{\rm sh}$ obtained with the help of Fig. 1 and after the fine fitting of the data belong to the range of barrier values found in other approaches, see Table 3. The values of the liquid-drop $B_{\rm f}^{\rm ld}$ and shell $B_{\rm f}^{\rm sh}$ contributions into the fission barriers of nuclei and the ground state $\beta_{\rm gs}$ and saddle point $\beta_{\rm sp}$ deformations of fissioning nuclei are near the corresponding values from Refs. \cite{kjs}. The values of the liquid-drop $B_{\rm f}^{\rm ld}$ contribution of the fission barrier are close to 10\% of the shell contribution $B_{\rm f}^{\rm sh}$.

\begin{table}
\caption{\label{tab4}The values of barrier of the fusion trajectory at the formation of the compound nucleus from the DNS $B_{\rm 0,CN}^{\rm DNS, f}$ and the parameter $\Delta_B$. The values of $\Delta_B$ and $B_{\rm 0,CN}^{\rm DNS, f}$ are given in MeV. The values of barriers $B_{\rm 0,CN}^{\rm DNS, f}$ is given relatively the ground state energy of the compound nucleus. The last column shows the laboratory, in which the experimental data are obtained for the reaction.}
\begin{center}
\begin{tabular}{cccc}
\hline
Reaction & $\Delta_B$ & $B_{\rm 0, CN}^{\rm DNS, f}$ & Exp. lab.\\
\hline
$^{208}$Pb+$^{50}$Ti & 5.0 & 12.5 & GSI, LBNL\\
$^{208}$Pb+$^{52}$Cr & 4.0 & 14.7 & LBNL\\
$^{208}$Pb+$^{54}$Cr & 11.0 & 11.1 & GSI\\
$^{208}$Pb+$^{58}$Fe & 7.0 & 10.6 & GSI, RIKEN\\
$^{208}$Pb+$^{64}$Ni & 6.5 & 6.2 & GSI,LBNL,RIKEN\\
$^{208}$Pb+$^{70}$Zn & 4.0 & 2.7 & GSI, RIKEN\\
$^{208}$Pb+$^{78}$Ge & 4.0 & 4.9 & \\
\hline
\end{tabular}
\end{center}
\end{table}

The values of damping parameter of the fission barrier $\gamma_D$ depend on the numbers of protons and neutrons in nucleus and belong to the range from $\approx 0.1$ to $0.03$ MeV$^{-1}$ \cite{snp,pnsk,pnsk-a}. The values of $\gamma_D$ presented in Table 3 belong to this range. The influence of this parameter on the evaporation characteristics are widely discussed, see, for details, Refs. \cite{dp,ds,ds2019,dh,snp,pnsk,pnsk-a,dns1,dns1c,dns2a,dns3,dns31,dns32,dns33,dns3a,dns4b,dns4c,dns4d,zg,param3}. Note that parameters $\gamma_D$ and $\gamma$ in Eq. (27) are different, because they are obtained by the fitting of different physical quantities.

As has pointed earlier, the accurate values of the collision energies of reaction leading to the SHN are not known due the thick target and the differences of the experimental data obtained in various laboratories. Therefore, it is reasonable to fit the maximum of the theoretical cross section using the parameter $\Delta_B$, because the position of the maximum depends strongly on it. We remind, that the parameter $\Delta_B$ is linked to the width of the fusion barrier distribution $g$, see Eqs. (4)-(5) and related text. The value of $g$ is not known experimentally. The value of $\Delta_B$ obtained from the fit of the data \cite{104lbnl} is given in Table 4. The value of $\Delta_B$ presented in Table corresponds to the value of the width of barrier distribution $g=3.3$ MeV, which belongs to the range of typical values of $g=2 - 4$ MeV used in other models \cite{zg,dns4b,dns4c,dns4d,dns5}. Note, the GSI data can be fitted by using larger values of $\Delta_B$.

The position of the barrier $B_{\rm 0,CN}^{\rm DNS, f}$ corresponds to the contacting near-spherical incident nuclei formed the DNS, see Fig. 1. Due to this the momentum of inertia related to this barrier, see Eq. (50), is calculated as $J_{\rm CN}^{\rm fus} \approx \mu (R_1+R_2)^2$.

\subsection{Reaction $^{208}$Pb($^{52}$Cr,$xn)^{260-x}$Sg}

\begin{figure}
\includegraphics[width=8.5cm]{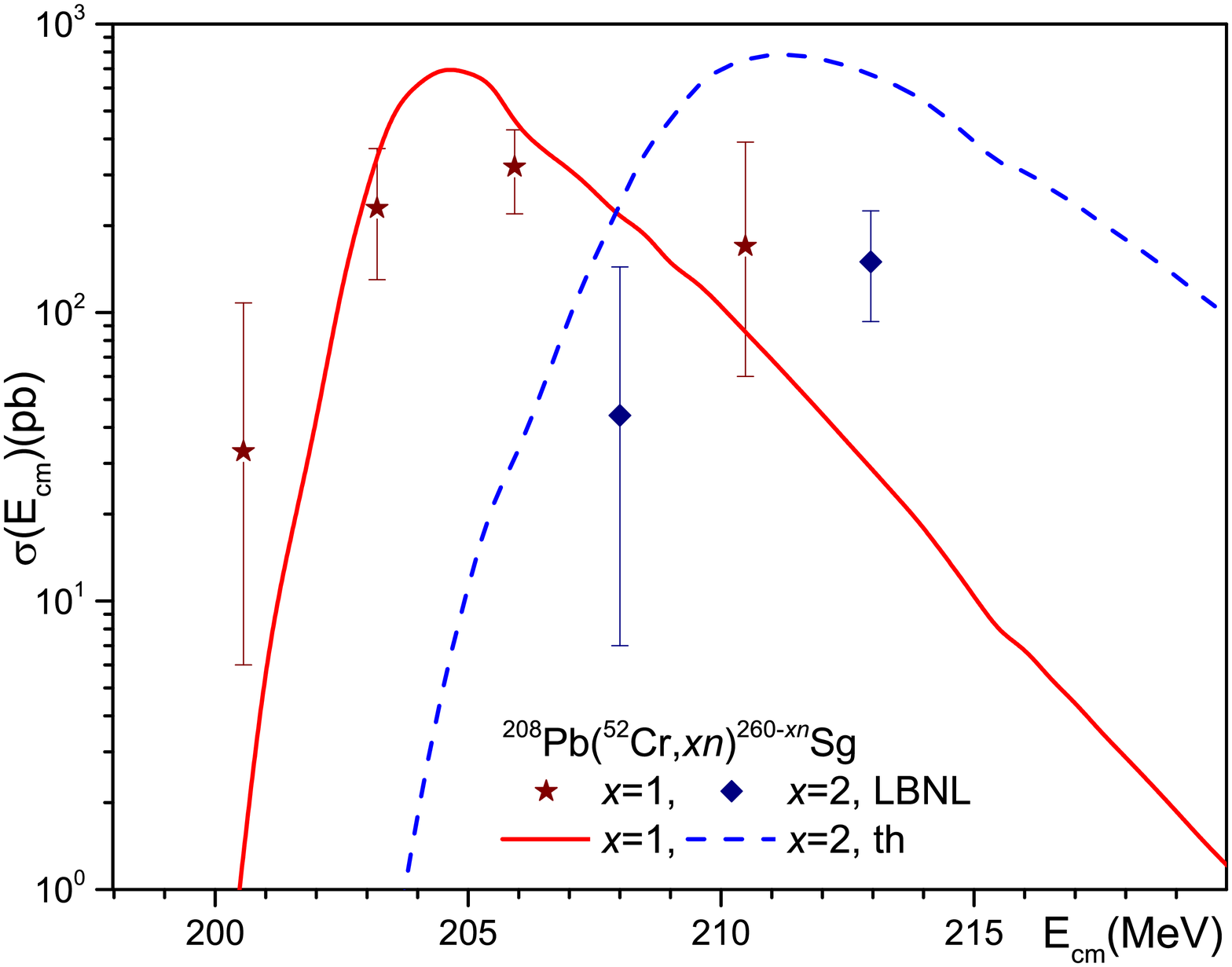}
\caption{\label{fig4} A comparison of our theoretical calculations of the cross sections for reactions $^{208}$Pb($^{52}$Cr,$xn)^{260-x}$Rf with $x=1$ and 2 with the experimental data measured in Ref. \cite{106lbnl} (LBNL).}
\end{figure}

The values of the cross sections for reactions $^{208}$Pb($^{52}$Cr,$xn)^{260-x}$Sg with $x=1$ and 2 are measured in LBNL \cite{106lbnl}. The results calculated in the framework of our model agree well with the experimental data, see Fig. 4. The parameters of the model used in the calculation of these reactions are presented in Tables 2--4. The values of parameters are chosen similar way as the ones for reactions $^{208}$Pb($^{50}$Ti,$xn)^{258-x}$Rf. The parameter values for reactions $^{208}$Pb($^{52}$Cr,$xn)^{260-x}$Sg are close to the corresponding ones for reaction $^{208}$Pb($^{50}$Ti,$xn)^{258-x}$Rf. However, we use a smaller value of the dissipation parameter $\gamma_D$, see Table 3, which is responsible for attenuation of the shell contribution of the fission barrier. Recall, that the value of $\gamma_D$ depends strongly on the numbers of protons and neutrons in nucleus \cite{snp,pnsk,pnsk-a}.

\subsection{Reaction $^{208}$Pb($^{54}$Cr,$xn)^{262-x}$Sg}

The cross sections for reactions $^{208}$Pb($^{54}$Cr,$xn)^{262-x}$Sg with $x=1$ and 2 are measured in GSI \cite{106gsi}. We calculate the cross sections for these reactions in the framework of our model. Our model describes well the experimental data, see Fig. 5.

The parameters of the model used in calculation of reactions $^{208}$Pb($^{54}$Cr,$xn)^{262-x}$Sg are given in Tables 2--4. The parameters presented in Table 4 are similar to the corresponding ones for reactions $^{208}$Pb($^{50}$Ti,$xn)^{258-x}$Rf and $^{208}$Pb($^{52}$Cr,$xn)^{260-x}$Sg.

\begin{figure}
\includegraphics[width=8.5cm]{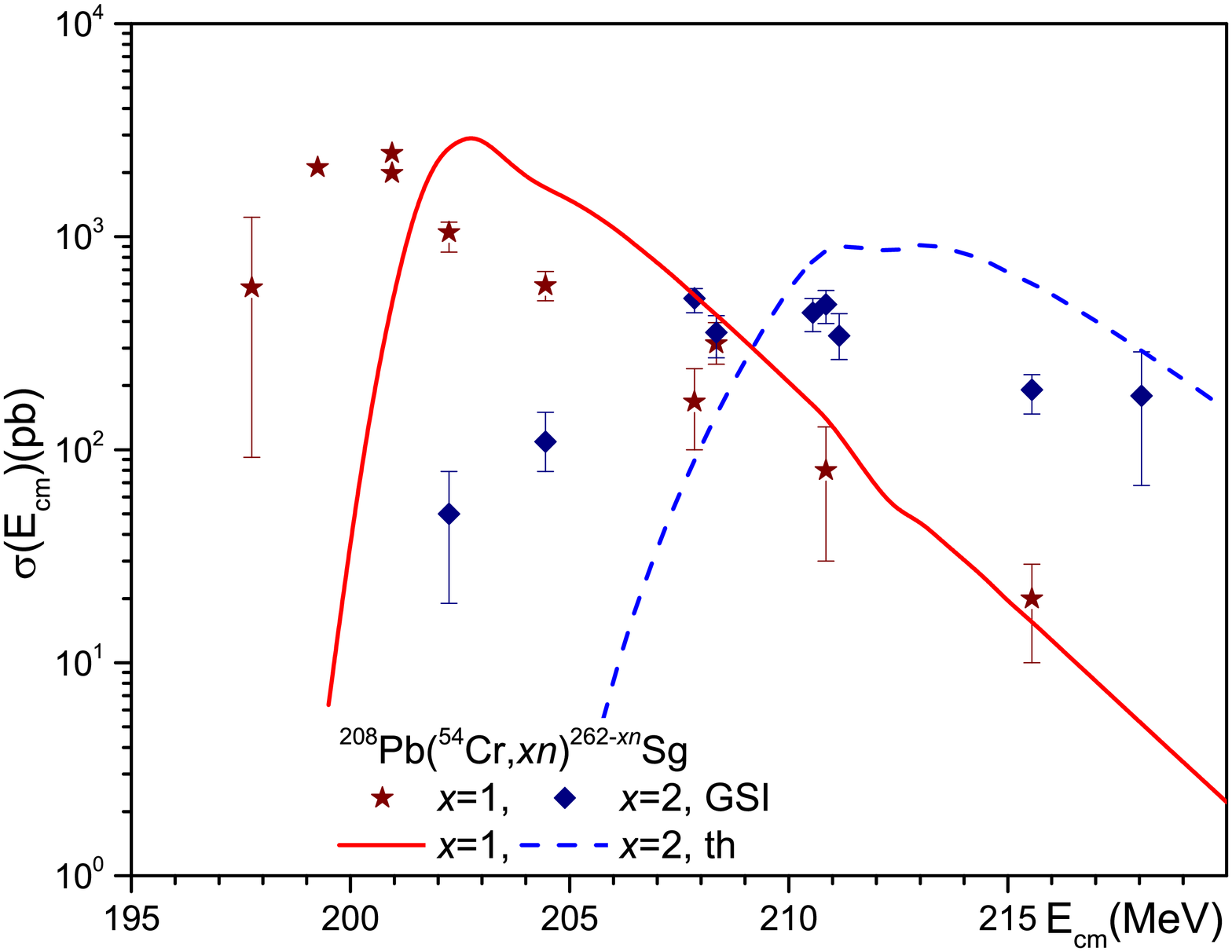}
\caption{\label{fig5} A comparison of our theoretical calculations of the cross sections for reactions $^{208}$Pb($^{54}$Cr,$xn)^{262-x}$Rf with $x=1$ and 2 with the experimental data measured in Ref. \cite{106gsi} (GSI).}
\end{figure}

We remind that the maxima of the SHN cross sections for reactions measured in GSI is taking place at smaller collision energies than the ones measured in other laboratories. Therefore, for the sake of the data description, we use the largest value of parameter $\Delta_B$ for this reaction in comparison to other reactions, see Table 4. The value of the barrier $B_{\rm 0,CN}^{\rm DNS, f}$ for $^{208}$Pb($^{52}$Cr,$xn)^{260-x}$Sg reaction is higher than the one for the reaction $^{208}$Pb($^{54}$Cr,$xn)^{262-x}$Sg, see Table 4. The excitation energy of the compound nucleus at the maximum of the SHN cross section reduces, when the maximum is shifted to smaller collision energies. Therefore, we change the parameters related to the survival probability of the compound nucleus. The values of fission barriers for these nuclei presented in Table 3 are close to the ones obtained in Refs. \cite{etfsi,mol}.

\subsection{Reaction $^{208}$Pb($^{58}$Fe,$xn)^{266-x}$Hs}

The values of the cross sections for reaction $^{208}$Pb($^{58}$Fe,$1n)^{265}$Hs are measured in GSI \cite{h,108gsi} and for reactions $^{208}$Pb($^{58}$Fe,$xn)^{266-x}$Hs with $x=1$ and 2 are obtained in RIKEN \cite{108riken}. The cross sections for these reactions calculated in the framework of our model agree well with the experimental data, see Fig. 6.

The difference in the collision energy values of the cross section maxima measured in GSI and RIKEN for reaction $^{208}$Pb($^{58}$Fe,$1n)^{265}$Hs is close to 4 MeV, see Fig. 6. It is reasonable to make the maximum of the cross section in our model for reaction $^{208}$Pb($^{58}$Fe,$1n)^{265}$Hs at the collision energy between the ones obtained in GSI and RIKEN. Due to this the value of $\Delta_B$ for this reaction lies between the ones for reactions with $^{52}$Cr and $^{54}$Cr projectiles, see Table 4. The width of the cross section peaks are close to the experimental ones.

\begin{figure}
\includegraphics[width=8.5cm]{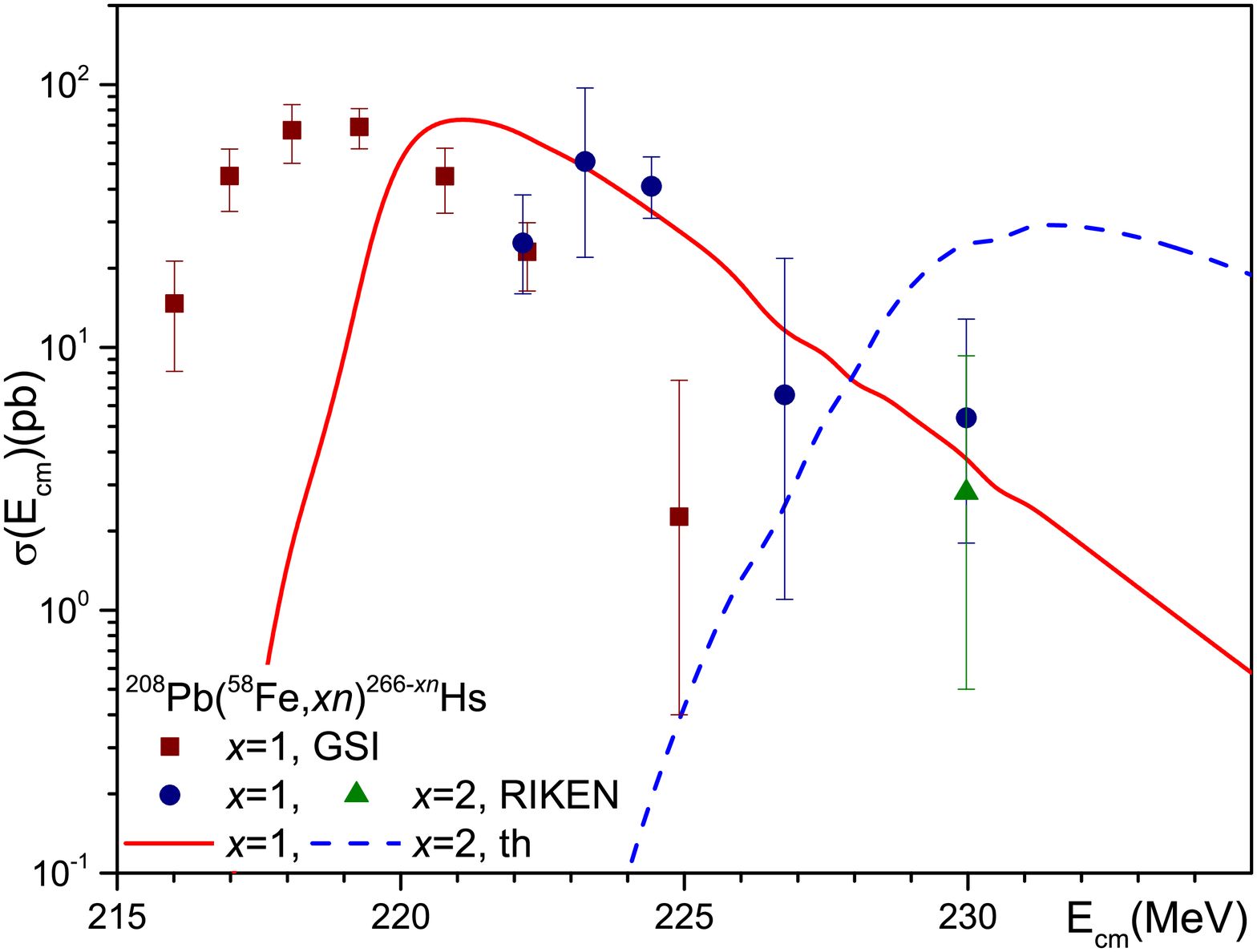}
\caption{\label{fig6} A comparison of our theoretical calculations of the cross sections for reactions $^{208}$Pb($^{58}$Fe,$xn)^{266-x}$Hs with $x=1$ and 2 with available experimental data. The cross sections for reaction $^{208}$Pb($^{58}$Fe,$1n)^{265}$Hs are measured in Refs. \cite{h,108gsi} (GSI) and for reactions $^{208}$Pb($^{58}$Fe,$xn)^{266-x}$Hs $x=1$ and 2 are obtained in Refs. \cite{108riken} (RIKEN).}
\end{figure}

The other parameters of the model used in the calculation of reactions $^{208}$Pb($^{58}$Fe,$xn)^{266-x}$Hs are presented in Tables 2--4. The values of parameters are chosen similar way as for other reactions.

Our value of the fission barrier for $^{266}$Hs is very close to the one obtained in Ref. \cite{mol} and larger the one presented in Ref. \cite{etfsi}. Our value of the fission barriers for $^{265}$Hs lies between the ones obtained in Refs. \cite{etfsi,mol}.

\subsection{Reaction $^{208}$Pb($^{64}$Ni,$xn)^{272-x}$Ds}

The cross sections for reaction $^{208}$Pb($^{64}$Ni,$1n)^{271}$Ds are measured in GSI \cite{h,108gsi}, LBNL \cite{110lbnl}, and RIKEN \cite{110riken}. We calculate the cross sections for reactions $^{208}$Pb($^{64}$Ni,$xn)^{272-x}$Ds with $x=1$ and $2$ in the framework of our model. The calculated values agree well with the available experimental data, see Fig. 7.

The parameters of the model used in the calculation of reactions $^{208}$Pb($^{64}$Ni,$1n)^{271}$Ds are given in Tables 2--4. We select the value of parameter $\Delta_B$ so that the theoretical peak of the cross sections lies close to the RIKEN data. This value of parameter $\Delta_B$ is close to the ones obtained for a description of the RIKEN data with other projectiles, see Table 4. The value of the barrier of the fusion trajectory from the DNS to the compound nucleus $B_{\rm 0 CN}^{\rm DNS, f}$ is estimated using Fig. 2. We see that the values $B_{\rm 0,CN}^{\rm DNS f}$ decrease with rising to projectile mass, see Figs. 1--2 and Table 4.
The smallest value of the dissipation parameter $\gamma_D$ is used for this reaction, see Table 3. The values of other parameters are chosen similar way as before.

The cross-section points of this reaction measured in GSI, see Fig. 7, locate at smaller collision energies, than the ones measured in RIKEN and the results obtained in our model. The difference in the position of the cross-section maxima for this reaction measured in different laboratories is close to 5 MeV.

\begin{figure}
\includegraphics[width=8.5cm]{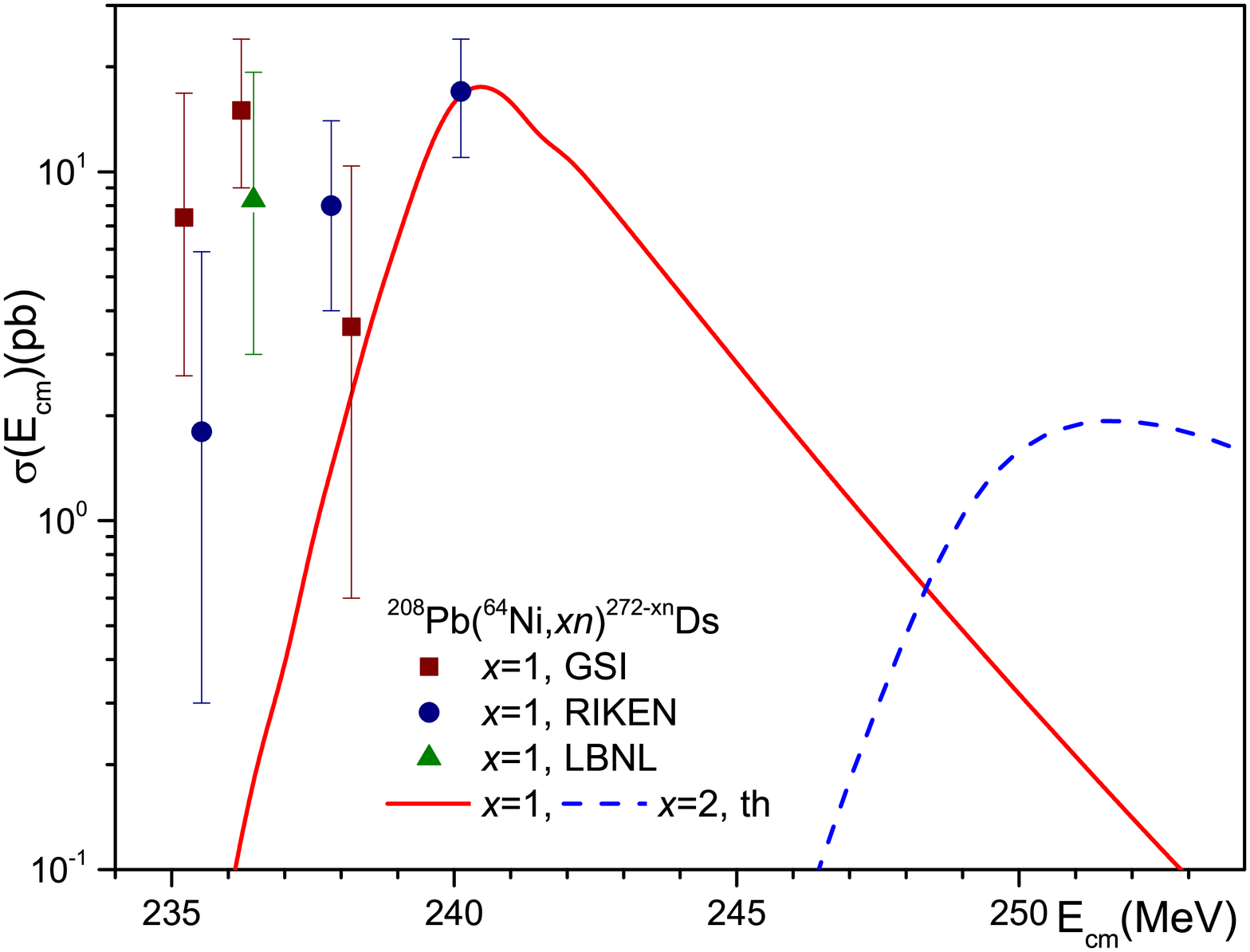}
\caption{\label{fig17} A comparison of our theoretical calculations of the cross sections for reactions $^{208}$Pb($^{64}$Ni,$1n)^{271}$Ds with available experimental data. The cross sections for this reaction are measured in Refs. \cite{h,108gsi} (GSI), \cite{110riken} (RIKEN) and \cite{110lbnl} (LBNL). The theoretical calculation of the cross section for reaction $^{208}$Pb($^{64}$Ni,$2n)^{270}$Ds is also presented.}
\end{figure}

The maximum of the cross section for the reaction with 2n emission $^{208}$Pb($^{64}$Ni,$2n)^{270}$Ds is located at high beam energy, which is not used in any experiments. The calculated value of cross section in the maximum for reaction $^{208}$Pb($^{64}$Ni,$2n)^{270}$Ds is strongly lower than the one for reaction $^{208}$Pb($^{64}$Ni,$1n)^{271}$Ds, see Fig. 7.

\subsection{Reaction $^{208}$Pb($^{70}$Zn,$xn)^{278-x}$Cn}

The cross sections for reaction $^{208}$Pb($^{70}$Zn,$1n)^{277}$Cn are measured in GSI \cite{h,112gsi} and RIKEN \cite{112riken,112riken1}. We calculate the cross sections for this reaction in the framework of our model. We describe well the data from both laboratories in our model by the corresponding parameter choice, see Fig. 8. The values of parameters used in our calculations of reactions $^{208}$Pb($^{70}$Zn,$xn)^{277-x}$Cn with x=1 and 2 are presented in Tables 2--4.

The value of barrier $B^{\rm DNS,tr}_{0,{\rm CN}}$ for this system obtained in our approach is 269.3 MeV, see Table 5. This value is close to the one evaluated in Ref. \cite{dns3a} 20.98 MeV - $Q_{\rm CN}$ = 20.98 MeV + 244.2 MeV = 265.18 MeV. As pointed in Ref. \cite{adamian}, the calculations of the nucleus-nucleus potentials of spherical nuclei in the framework of DNS model have the Coulomb barriers which are at least 5 MeV lower than the phenomenological Bass barriers \cite{bass}. The values of barriers of the nucleus-nucleus potential for very asymmetric systems with medium or small values of $Z_1 Z_2$ calculated in our approach are close to the Bass barriers \cite{d2015}. Therefore, such difference between values of the barrier $B^{\rm DNS,tr}_{0,{\rm CN}}$ obtained in different approaches is reasonable.

\begin{figure}
\includegraphics[width=8.5cm]{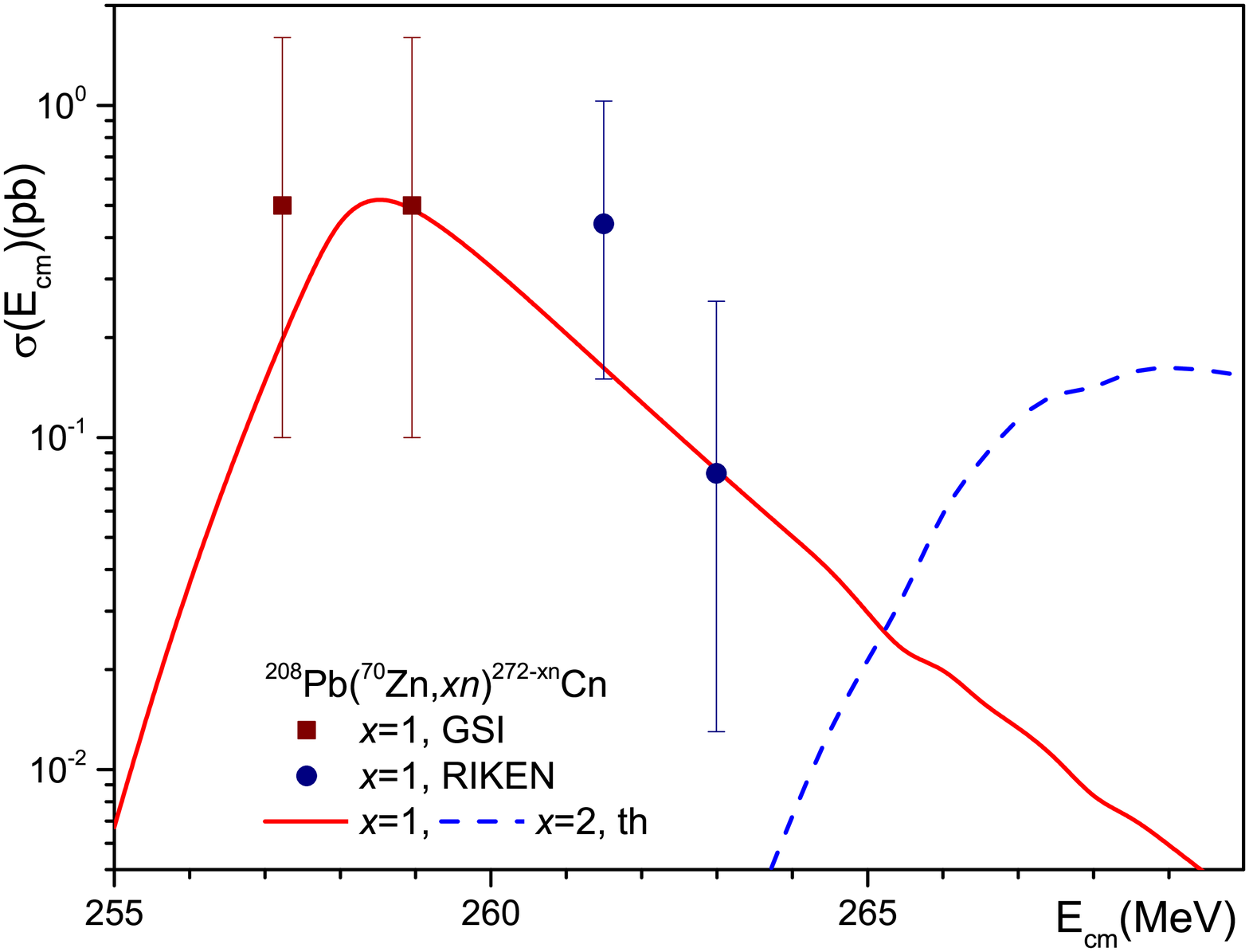}
\caption{\label{fig8} A comparison of our theoretical calculation of the cross section for reactions $^{208}$Pb($^{70}$Zn,$1n)^{277}$Cn with available experimental data. The cross sections for this reaction are measured in Refs. \cite{h,112gsi} (GSI) and \cite{112riken,112riken1} (RIKEN). The theoretical calculation of the cross section for reaction $^{208}$Pb($^{70}$Zn,$2n)^{276}$Cn is also presented.}
\end{figure}

The mechanism of this reaction is taken into account the contribution of the intermediate state. This intermediate state locates in the well between the inner and outer fission barriers, see Fig. 2. The height of the inner fission barrier is lower the height of the outer fission barrier on approximately 1 MeV, see Fig. 2. The height of barrier related to the decay back to the DNS is higher than the heights of the inner fission barriers, see Fig. 2 and Tables 3,4. Therefore, the decay of the intermediate state to the compound nucleus is the most preferable for this reaction. Due to this we put $P_{is} \approx 0.88(0.85)$ for $1n(2n)$ reactions for a simplification of the calculation.

\subsection{Reaction $^{208}$Pb($^{78}$Ge,$xn)^{286-x}$Fl}

The values of the cross sections for reaction $^{208}$Pb($^{78}$Ge,$xn)^{285}$Fl for $x=1$ and 2 have not been measured up to now. We calculate the cross sections for these reactions in the framework of our model and present it in Fig. 9. The values of SHN production cross section are very small for these reactions.

\begin{figure}
\includegraphics[width=8.5cm]{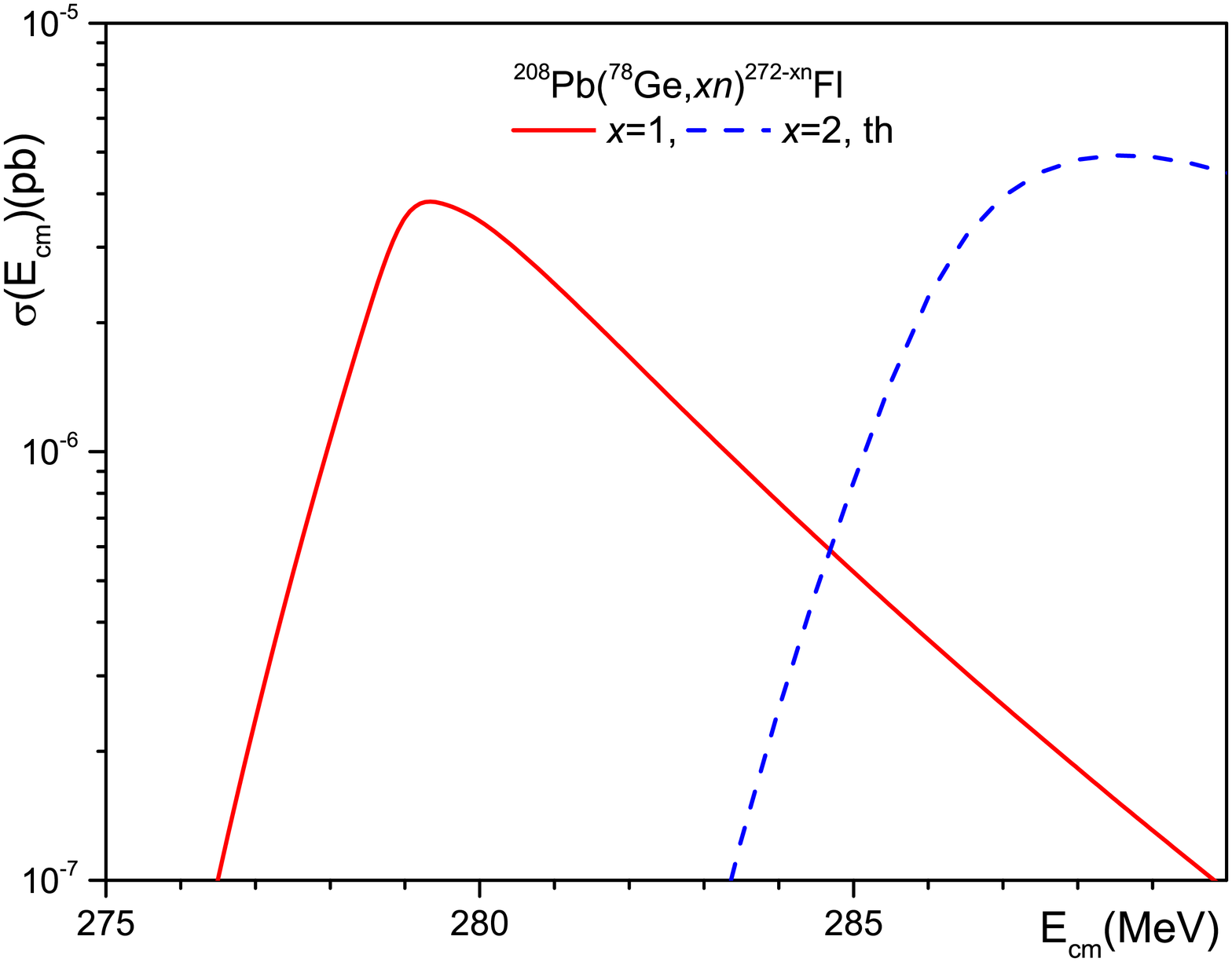}
\caption{\label{fig9} The results of calculations of the cross sections for reactions $^{208}$Pb($^{78}$Ge,$xn)^{286-x}$Fl for $x=1$ and 2.}
\end{figure}

The parameters of the model used in the calculation of reactions $^{208}$Pb($^{78}$Ge,$xn)^{285}$Fl are given in Tables 2--4. The values of parameters are chosen using a similar procedure as before.

The mechanism of compound nucleus formation in this reaction is also passed through the intermediate state, see Fig. 2. The height of the inner fission barrier is slightly higher than the height of the outer fission barrier, see Fig. 2. The height of barrier related to the decay back to the DNS is higher than the ones of the inner or outer fission barriers, see Fig. 2 and Tables 3,4. Therefore, the decay of the intermediate state into the quasi-fission fragments is more preferable than decays into the compound nucleus or the DNS. Due to this we put $P_{is} \approx 0.4$ for a simplification. The decay of the intermediate state into the quasi-fission fragments for reaction $^{208}$Pb($^{78}$Ge,$xn)^{286-x}$Fl is more probable in a comparison to reaction $^{208}$Pb($^{70}$Zn,$xn)^{278-x}$Cn.

\subsection{The probability of compound nucleus formation}

Let us discuss the mechanisms of compound nucleus formation in our model by comparing the barrier heights of different DNS decay processes. The various barrier heights are given in Tables 1 and 5.

The barrier heights related to the transfer $B^{\rm DNS,tr}_{0,{\rm CN}}$ and fusion $b_{0,{\rm CN}}^{\rm DNS, f}=B^{\rm DNS,f}_{0,{\rm CN}}-Q_{\rm CN}$ paths of the compound nucleus formation obey the inequality $B^{\rm DNS,tr}_{0,{\rm CN}}>b_{\rm 0,CN}^{\rm DNS, f}$. We can conclude from this inequality that $\Gamma_{\rm CN}^{\rm DNS, tr}(E,\ell)=0$ at small collision energies and $ \Gamma_{\rm CN}^{\rm DNS, f}(E,\ell) \gg \Gamma_{\rm CN}^{\rm DNS, tr}(E,\ell)$ at high collision energies. Therefore, the transfer mechanism of compound nucleus formation has a small contribution at high collision energies only.

\begin{table}
\caption{\label{tab5}The values of barrier heights for the DNS decay branches. $B^{\rm fus}_0$ is the barrier for the DNS decay into the incident channel, $b_{\rm 0,CN}^{\rm DNS, f}$ is the barrier for the DNS decay into the compound nucleus by the fusion pass, $B^{\rm DNS,tr}_{0,{\rm CN}}$ is the barrier for the DNS decay into the compound nucleus by the nucleon transfer way, $B^{\rm DNS}_{0,{\rm DIC}}$ is the barrier for the DNS decay into the more symmetric nuclear system than the incident channel, and $B^{\rm DNS}_{0,{\rm def}}$ is the barrier for the DNS decay into the incident deformed nuclei. $\delta_B=b_{\rm 0,CN}^{\rm DNS, f}-B^{\rm DNS}_{0,{\rm def}}$. The barriers are evaluated relatively the nucleus-nucleus interaction energy on the infinite distance between them at $\ell=0$. All values are given in MeVs.}
\begin{center}
\begin{tabular}{ccccccc}
\hline
Reactions & $B^{\rm fus}_0$ & $b_{\rm 0, CN}^{\rm DNS, f}$ & $B^{\rm DNS,tr}_{0,{\rm CN}}$ & $B^{\rm DNS}_{0,{\rm DIC}}$ & $B^{\rm DNS}_{0,{\rm def}}$ & $\delta_B$ \\
\hline
$^{50}$Ti + $^{208}$Pb & 189.8 & 182.0 & 198.7 & 187.1 & 175.8 & 6.2 \\
$^{52}$Cr + $^{208}$Pb & 207.2 & 198.4 & 212.3 & 199.1 & 191.3 & 7.1 \\
$^{54}$Cr + $^{208}$Pb & 206.0 & 198.2 & 215.7 & 202.1 & 189.1 & 9.1 \\
$^{58}$Fe + $^{208}$Pb & 222.0 & 215.6 & 232.6 & 217.5 & 204.0 & 11.6 \\
$^{64}$Ni + $^{208}$Pb & 236.6 & 231.1 & 251.9 & 233.9 & 216.2 & 14.9 \\
$^{70}$Zn + $^{208}$Pb & 251.1 & 246.9 & 269.3 & 248.0 & 228.4 & 18.5 \\
$^{78}$Ge + $^{208}$Pb & 264.4 & 270.3 & 289.9 & 265.4 & 239.9 & 30.4 \\
\hline
\end{tabular}
\end{center}
\end{table}

We see in Table 5 that for each reaction the height of barrier $B^{\rm DNS}_{0,{\rm def}}$ are the lowest and $B^{\rm DNS,tr}_{0,{\rm CN}} > b_{\rm 0,CN}^{\rm DNS, f}$. Therefore, the values of the widths obey the inequalities: $\Gamma_{\rm def}^{\rm DNS}(E,\ell) \gg \Gamma_{\rm CN}^{\rm DNS, tr}(E,\ell)$, $\Gamma_{\rm def}^{\rm DNS}(E,\ell) \gg \Gamma_{\rm DIC}^{\rm DNS}(E,\ell)$, and $\Gamma_{\rm def}^{\rm DNS}(E,\ell) \gg \Gamma_{\rm sph}^{\rm DNS}(E,\ell)$. As a result, $\Gamma^{\rm tot}_{\rm CN}(E,\ell) \approx \Gamma_{\rm def}^{\rm DNS}(E,\ell)$ and the probability of compound nucleus formation may be approximated as
\begin{eqnarray}
P(E,\ell) \approx \frac{\Gamma_{\rm CN}^{\rm DNS, f}(E,\ell)}{\Gamma_{\rm def}^{\rm DNS}(E,\ell)} .
\end{eqnarray}
At the high excitation energy $E^*=E+Q_{\rm CN}$ of the compound nucleus this expression can be presented in a simple form
\begin{eqnarray}
P(E^*,\ell) &\approx& \frac{\Gamma_{\rm CN}^{\rm DNS, f}(E^*,\ell)}{\Gamma_{\rm def}^{\rm DNS}(E^*,\ell)} \nonumber \\ &\propto & \frac{\exp{ \left\{ 2 \left[ a_{\rm dens} \left(E^*-B_{0,{\rm CN}}^{\rm DNS, f} \right) \right]^{1/2} \right\} }}{ \exp{ \left\{ 2 \left[ a_{\rm dens} \left( E^*-(B_{0,{\rm def}}^{\rm DNS}+Q_{\rm CN}) \right) \right]^{1/2} \right\} }} \nonumber \\ &\approx& \exp{ \left\{-\left[ \frac{a_{\rm dens}}{E^*}\right]^{1/2} \delta_B \right\} },
\end{eqnarray}
where $\delta_B=b_{\rm 0, CN}^{\rm DNS, f}-B^{\rm DNS}_{0,{\rm def}}$.

The difference between barriers $\delta_B$ increases from 6.2 MeV for reaction $^{50}$Ti + $^{208}$Pb to 30.4 MeV for reaction $^{78}$Ge + $^{208}$Pb, see Table 5. Due to this and Eq. (57), the probability of compound nucleus formation $P(E^*,\ell)$ strongly decreases with the increasing of the mass of the projectile, see also Fig. 10. The dependencies of the probabilities of compound nucleus formation $P(E^*,\ell)$ on the excitation energy of the compound nucleus formed in reactions $^{50}$Ti, $^{52,54}$Cr, $^{58}$Fe, $^{64}$Ni, $^{70}$Zn, and $^{78}$Ge + $^{208}$Pb for $\ell=0$ are presented in Fig. 10.

According to the statistical approach, see Eq. (57), the probability of compound nucleus formation increases with the rising of energy $E^*$. This rising is very strong at the energies near the barrier of the compound nucleus formation by fusion trajectory pass. This tendency is clearly seen in Fig. 10. The probability of compound nucleus formation rises smoother at higher energies.

\begin{figure}
\includegraphics[width=8.5cm]{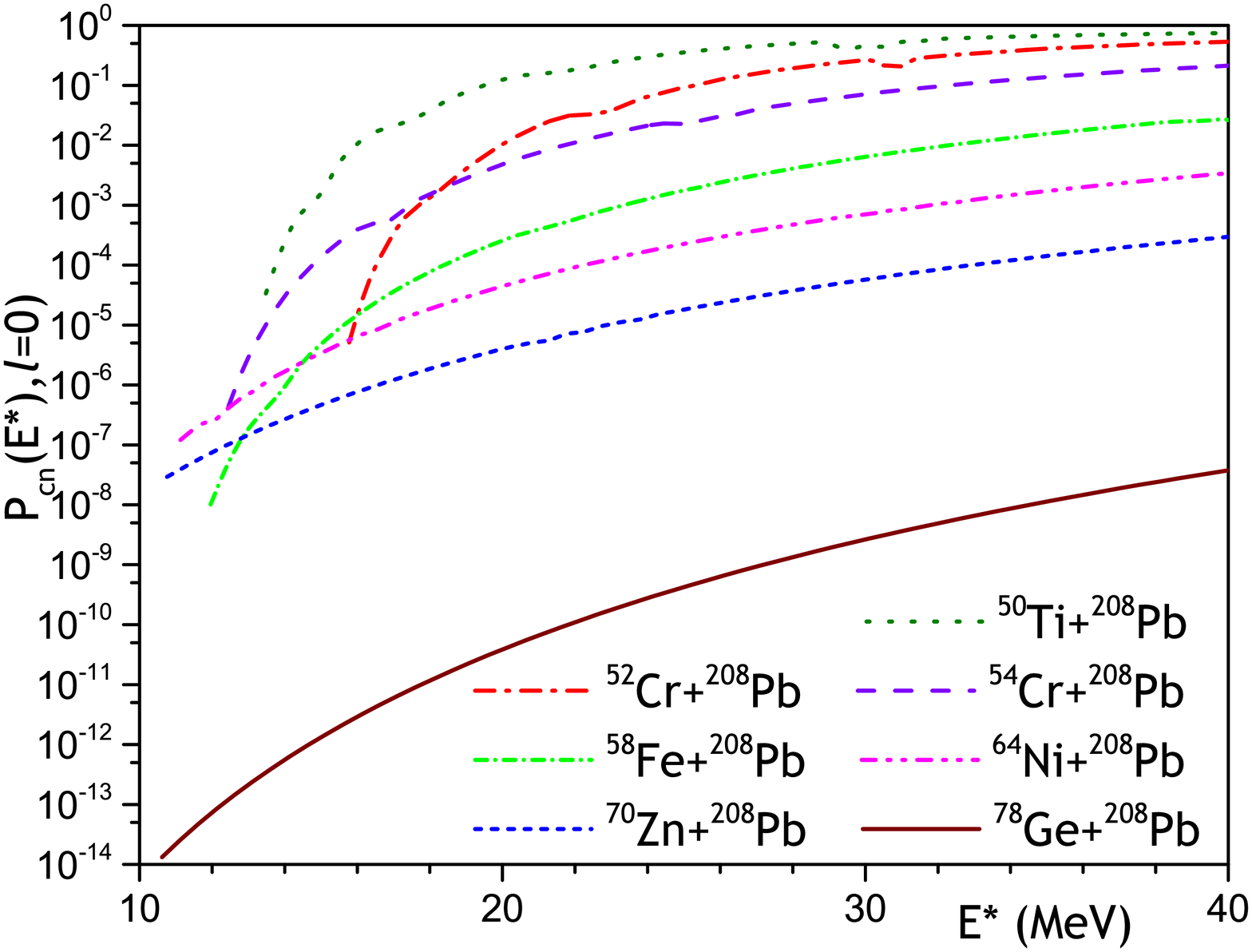}
\caption{\label{fig10} The dependencies of the probability of compound nucleus formation on the excitation energy of the compound nucleus formed in reactions $^{50}$Ti, $^{52,54}$Cr, $^{58}$Fe, $^{64}$Ni, $^{70}$Zn, and $^{78}$Ge + $^{208}$Pb for $\ell=0$.}
\end{figure}

The value of probability $P(E^*,\ell)$ for reaction $^{78}$Ge + $^{208}$Pb is strongly smaller than the one for reaction $^{70}$Zn + $^{208}$Pb, see Fig. 10. This is related to the rise of the difference between barriers $\delta_B$ for reaction with $^{78}$Ge projectile, see Table 5. Due to this the cross sections for reaction $^{78}$Ge + $^{208}$Pb is strongly smaller than the one for reaction $^{70}$Zn + $^{208}$Pb, compare results presented in Figs. 9 and 10.

We remind that the values of the barrier height between deformed incident nuclei $B^{\rm DNS,f}_{0,{\rm def}}$ as well as the width $\Gamma_{\rm def}^{\rm DNS}(E,\ell)$ depend strongly on the surface stiffness coefficient. Therefore, the value of the surface stiffness coefficient affects considerably on the probability of the compound nucleus formation. Due to this, the choice of incident nuclei is very important for the values of compound nucleus production cross section.

The crucial role of the transfer mechanism of the compound nucleus formation in the framework of the DNS model is related to the small height of the barrier $B^{\rm DNS,tr}_{0,{\rm CN}}$ in this model, see discussion of this barrier height in Sec. III.F. The DNS path of compound nucleus formation is main, when $b^{\rm DNS,f}_{0,{\rm CN}}>B^{\rm DNS,tr}_{0,{\rm CN}}$. In our model $b^{\rm DNS,f}_{0,{\rm CN}}<B^{\rm DNS,tr}_{0,{\rm CN}}$, therefore the fusion path of the compound nucleus production is basic. So, the role of different mechanisms of SHN production depends the potential landscape, which is defined by the model(s) for the calculation of the potential energy for one- and two-body nuclear shapes.

\section{Conclusion}

We have presented the new model of the SHN production in the cold-fusion reactions. This model takes into account the competition between the DNS multi-nucleon transfer and fusion trajectories of the compound nucleus formation. The available experimental data are well described in our model.

The compound nucleus is mainly formed by the fusion path, because the barrier related to this fusion way is lower than the barrier related to the multi-nucleon transfer from the light nucleus to heavy one (the DNS trajectory).

We show the correlation between the surface stiffnesses of nuclei involved in the SHN production and the reaction cross sections. The using of more stiff nuclei leads to higher cross section value due to reduction of the DNS decay width to deformed nuclei.

The competition between the compound nucleus formation and the true quasi-fission occurred along the fusion trajectory is taken into account for heavy nucleus-nucleus systems leading to the SHN. This competition is related to the existence of the intermediate state and connected to the landscape of the potential energy surface. The intermediate state is important for reactions with heavy projectiles. The quasi-fission is linked to the decay of the intermediate state into fragments bypassing the formation of the compound nucleus.

The yields of the various reaction processes in the model depend on the relative values of the corresponding barrier heights. The values of the barrier heights $B_{\rm 0, CN}^{\rm DNS, f}$ and $B_{\rm f}$ depend on the choice of the nuclear structure model for calculation of these barriers and the one-body shape parametrization. The uncertainty of the fission barrier height $B_{\rm f}$ obtained in the framework of different models is several MeVs, see Tabl. 3. So, we may expect similar uncertainty for the barrier $B_{\rm 0, CN}^{\rm DNS, f}$. The values of the barriers $B^{\rm fus}_0$, $B^{\rm DNS,tr}_{0,{\rm CN}}$, $B^{\rm DNS}_{0,{\rm DIC}}$ and $B^{\rm DNS}_{0,{\rm def}}$ are determined by the choice of the nuclear part of the nucleus-nucleus potential and the stiffness parameter of the nuclei. The difference between the interaction barrier heights of spherical nuclei leading to SHNs calculated in various approaches to the nuclear interaction part may reach 20 MeV \cite{dn}. We emphasize the barrier values calculated by using the potential (6) agree well with the ones, extracted from the quasi-elastic scattering, see Table 1, and with the empirical barriers for light and medium heavy-ion systems \cite{d2015}. The values of stiffness parameter are only known for some nuclei. Thus, the obtained results are model dependent, and the using more accurate approaches are encouraged for further studies.

\section*{Acknowledge}

V. Yu. Denisov is grateful to Academician Yu. Ts. Oganessian and Professor S. Hofmann for the stimulated discussions. V. Yu. Denisov is grateful to Professor S. Hofmann for a digital form of the experimental data for reactions $^{208}$Pb($^{58}$Fe,$1n)^{265}$Hs and $^{208}$Pb($^{64}$Ni,$1n)^{271}$Ds \cite{108gsi}, which were presented in Ref. \cite{dh} too.

\end{document}